\documentclass[12pt]{iopart}

\expandafter\let\csname equation*\endcsname\relax
\expandafter\let\csname endequation*\endcsname\relax

\usepackage{xcolor}
\usepackage{graphicx}
\usepackage[font = footnotesize, labelfont = bf]{caption}
\usepackage{subcaption}
\usepackage{float}
\usepackage{mathtools}
\usepackage{bbold}

\usepackage{physics}
\usepackage{tikz}
\usepackage{tikz-3dplot}
\usepackage{pgfplots}
\usetikzlibrary{shapes}
\tdplotsetmaincoords{60}{110}
\pgfplotsset{compat=1.14}

\usetikzlibrary{arrows,decorations.markings}

\newcommand{\change}[1]{#1} 

\newcommand{\heff}{\hbar_{\text{eff}}}
\newcommand{\hbarE}{\ensuremath{\hbar_{\rm eff}}}

\newcommand{\ha}[1]{\ensuremath{\hat{a}_{#1}}}
\newcommand{\had}[1]{\ensuremath{\hat{a}^{\dagger}_{#1}}}
\newcommand{\hH}{\hat{H}}

\renewcommand{\d}{\ensuremath{\operatorname{d}\!}}

\newcommand{\coh}[1]{\ensuremath{\vert{#1}\rangle_{\rm coh}}}

\newcommand{\proj}[1]{\ensuremath{|{#1}\rangle_{\rm proj}^{N}}}

\newcommand{\vphi}{\ensuremath{\Vec{\phi}}}

\newcommand{\vad}{\ensuremath{\hat{\Vec{a}}^{\dagger}}}

\usepackage{iopams}  
\newcommand{\RegensburgUniversity}{Institut f\"ur Theoretische Physik, Universit\"at Regensburg, D-93040 Regensburg, Germany}
\newcommand{\Steve}{Department of Physics and Astronomy, Washington State University, Pullman, WA USA}

\begin{document}

\title{Controlling Many-Body Quantum Chaos: Bose-Hubbard systems}

\author{Lukas Beringer}
\address{\RegensburgUniversity}
\author{Mathias Steinhuber}
\address{\RegensburgUniversity}
\author{Juan Diego Urbina}
\address{\RegensburgUniversity}
\author{Klaus Richter}
\address{\RegensburgUniversity}
\author{Steven Tomsovic}
\address{\RegensburgUniversity}
\address{\Steve}

\ead{\change{Lukas.Beringer@physik.uni-regensburg.de}}

\begin{abstract}
This work develops a quantum control application of many-body quantum chaos for ultracold bosonic gases trapped in optical lattices. It is long known how to harness exponential sensitivity to changes in initial conditions for control purposes in classically chaotic systems.  In the technique known as \textit{targeting}, instead of a hindrance to control, the instability becomes a resource.  Recently, this classical targeting has been generalized to quantum systems either by periodically countering the inevitable quantum state spreading or by introducing a control Hamiltonian, where both enable localized states to be guided along special chaotic trajectories toward any of a broad variety of desired target states. Only strictly unitary dynamics are involved; i.e., it gives a \textit{coherent quantum targeting}.  In this paper, the introduction of a control Hamiltonian is applied to Bose-Hubbard systems in chaotic dynamical regimes.  Properly selected unstable mean field solutions can be followed \change{particularly} rapidly to states possessing precise phase relationships and occupancies.  In essence, the method generates a quantum simulation technique that can access rather special states.  The protocol reduces to a time-dependent control of the chemical potentials, opening up the possibility for application in optical lattice experiments. Explicit applications to custom state preparation and stabilization of quantum many-body scars are presented in one- and two-dimensional lattices (three-dimensional applications are similarly possible).  
\end{abstract}

%
%

\section{Introduction}
\label{sec:intro}

The exponential instability inherent in chaotic dynamics typically leads a system towards relaxation, equilibration, and/or thermalization.  Naturally, chaos poses fundamental challenges for controlling system dynamics as well.  Nevertheless, there are circumstances in which chaos provides a resource for control, such as \textit{gravity assist} conjectured by Ulam~\cite{Ulam58}, nicely illustrated as a control problem by the diversion of the International Sun-Earth Explorer satellite to a Giacobini-Zinner comet flyby~\cite{Farquhar85, Farquhar01}, and more comprehensively, by the theory of \textit{classical targeting}~\cite{Ott90, Shinbrot90, Kostelich93, Bollt95, Schroer97}, which falls under the more general moniker, \textit{controlling chaos}~\cite{Ott06}.  Roughly speaking, for fully chaotic systems the origin of the resource is the existence of heteroclinic motion (trajectories)~\cite{Poincare99}.  In a nutshell, there exists a set of neighboring initial conditions, which lead to trajectories that deviate with a maximal exponential divergence from the trajectory of the given initial conditions.  Likewise, but the inverse process, there exists a set of trajectories which approach the final conditions in a maximal exponential sense ending up in its neighborhood.  Those rare trajectories belonging to both sets capture the essence of heteroclinic motion, and give rise to the opportunity to make a tiny, but very specific, change in the initial conditions that accelerates greatly how quickly the system dynamically evolves to the predetermined final point.  Hence, a great deal of the effort in controlling classical chaos (targeting) is the identification of the precise and quite rare (heteroclinic) trajectory which accomplishes the goal in an optimal way.

\begin{figure}[ht]
	\centering
	\includegraphics[width = 0.8\textwidth]{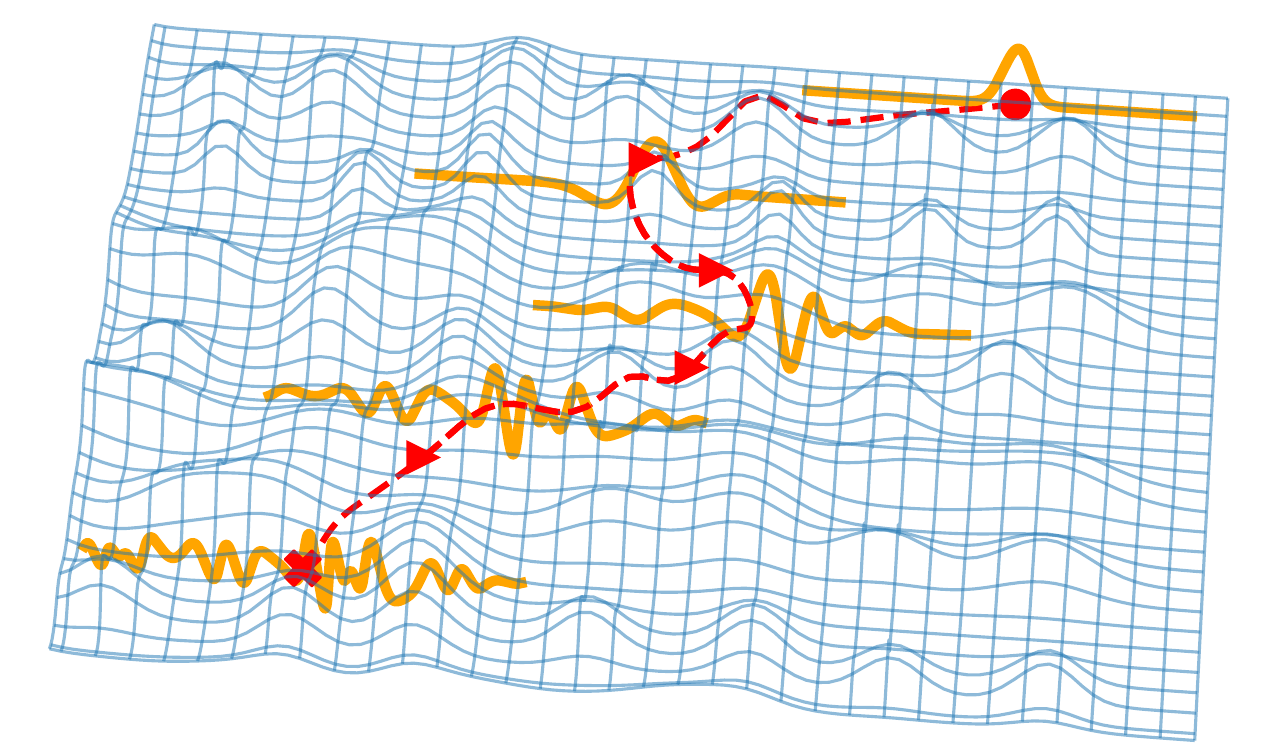}
	\caption{\textbf{Schematic illustration of an initially localized quantum state guided along a classical transport trajectory.}  A localized quantum state, here represented as an orange cut through its Wigner function, centered around an initial phase space point (upper right dot)
		can be guided to a specific target state (lower left cross) by following a solution of the classical limit. In a generic system exhibiting quantum chaos the localized state would equilibrate in a logarithmically short time period, and thus the spreading must be counteracted in a control protocol.
		The blue grid schematically represents a Hamiltonian landscape generating unstable dynamics.
	}
	\label{fig:quantum_control}
\end{figure}

Recently, the translation of classical targeting into the quantum realm has been introduced and applied to a longstanding paradigm of simple chaotic systems~\cite{Tomsovic23, Tomsovic23b}, the kicked rotor~\cite{Chirikov79, Izrailev90}.  For quantum systems with a well defined classical analog, it turns out that the same heteroclinic motion can be exploited both to arrive at a predetermined, possibly exotic, target or final state, and greatly accelerate the way there.  It can also be done with exclusively unitary dynamics, and thus be fully coherent.  The crucial new element not found in the classical case essentially arises from the uncertainty principle~\cite{Heisenberg27}.  The closest a quantum state can get to the initial conditions of a classical analog is in the form of a minimum uncertainty wave packet and a coherent state, depending on the context.  By virtue of a Wigner transform analysis~\cite{Wigner32} of these states, they correspond most closely to a Liouvillian density of neighboring initial conditions as opposed to a single trajectory.  Thus, a quantum analog of classical targeting must also control the Liouvillian flow around the heteroclinic trajectory; see Fig.~\ref{fig:quantum_control}. It is this extra requirement 
that led to the introduction of two methods of \textit{optimal coherent quantum targeting}.  The first technique follows the local stability analysis and counteracts from time to time the inevitable spreading of the density~\cite{Tomsovic23}.  The second technique is best conceived of as a form of quantum simulation~\cite{Johnson14} in which a quite different Hamiltonian simulates the heteroclinic evolution in a stable way~\cite{Tomsovic23b}.  In this work, the second technique is being applied to an important, experimentally realizable many-body system, i.e., the Bose-Hubbard model~\cite{Gersch63, Jaksch98}.

Quantum control techniques comprise a large longstanding research field with much of the early work prompted by the dream of controlling chemical reactions~\cite{Tannor85, Brumer86, Judson92, Warren93, Peirce88, Kosloff89}, \change{which more recently have been applied to many-body systems~\cite{Doria11,Frank16}}.  A survey~\cite{Dong10} and an overview of optimal control theory~\cite{James21} provide more information on this larger field.  On the whole, this field is not directly aimed at the challenge posed by quantum chaos, where exponential instability is converted from hindrance to resource.  Nevertheless, there are a few works~\cite{Tomsovic97, Gong05, Gruebele07, Bitter17, Vanhaele21, Vanhaele22, Madronero06}, but none address general approaches to optimal coherent targeting in quantum chaotic many-body systems. 

As alluded to before, a quantum targeting in an isolated many-body dynamical system may offer some important capabilities.  A fully chaotic dynamics eventually visits all possible states of a system independent of starting point, hence there is the possibility of placing a system in an otherwise quite unusual or difficult-to-access state.  In a space of enormous volume as a many-body system would possess, in most cases it would take an exceedingly long time to arrive at a desired state through an ergodic wandering, however, heteroclinic motion provides those rare, most rapid paths from particular starting to final states, thus providing a tremendously accelerated route to a desired state.

Many-body systems based on bosonic ultracold atoms provide ideal candidates for implementation of coherent quantum targeting for a number of reasons.  In the main though, there is a broad variety of experiments with great control over tunable parameters being performed, and they provide an excellent platform for quantum simulations~\cite{Bloch08, Dalibard11, Bloch12, Chien15, Langen15, Schafer20, Yang20, Altman21, Braun23, Bluvstein23}.  However, in addition they possess well defined classical analogs in the form of a mean-field limit with large particle number as required by the control protocol. The classical dynamics is governed by the Gross-Pitaevskii equation or its discretized version on a lattice (in optical lattices the system gets effectively described by a Bose-Hubbard model~\cite{Jaksch98}); see~\cite{Pitaevskii03} for a detailed review.  This allows for a phase space formulation in which strongly chaotic regimes can be identified~\cite{Kolovsky04,Pausch21} and where the resulting heteroclinic dynamics can be explored.  Finally, the systems admit both a continuous $U(1)$ and discrete dynamical symmetries that can be taken advantage of in developing control protocols for the purposes of introducing extremely helpful mappings between different dimensional lattices and expanding the range of initial states to which the protocol can be applied. For this work it is critical to be able to switch off the on-site interaction between the atoms \cite{Chin10, Su23, Impertro23}, and control the chemical potentials of the sites.

This paper is structured as follows: in the next section background material and notation are introduced related to  dynamical considerations, aspects of heteroclinic motion, Bose-Hubbard models, coherent and Fock states resulting from number-projected coherent states, truncated Wigner approximations, and finally control protocol considerations.  This is followed in Sec.~\ref{sec:1d_applications} with coherent quantum targeting applied to one-dimensional, $1D$, lattices.  The realization of the protocol reduces to switching off the interactions and controlling the chemical potentials of the individual sites in a time-dependent fashion. Examples of state preparation with non-trivial condensate phases as well as periodic many-body states are given. Error sources and fidelity of the protocol is discussed for explicit examples. Section~\ref{sec:2d_applications} gives the application to two-dimensional, $2D$, lattices.  An example of a lattice of discrete vortices is given in addition to periodic mean field solutions.  Finally, there is a summary and outlook for further research.

\section{Background}
\label{sec:background}

In order to describe coherent quantum targeting and its application to Bose-Hubbard models, it is helpful to start with some common background concerning chaotic dynamics, stability analysis of dynamical systems, Bose-Hubbard systems, their mean field limit and symmetries, coherent and number projected coherent states, truncated Wigner approximations, and the protocol for the optimal coherent quantum targeting itself.

\subsection{Chaos, heteroclinic motion, and times scales}
\label{sec:chaos}

Within a chaotic volume of phase space, given any arbitrary initial and target state that respects the restrictions due to constants of the motion, ergodicity guarantees the existence of a transport pathway connecting the respective states.  In fact, almost all of the trajectories would provide a connection.  However, the time scale for making the connection with a typical ergodic trajectory would be prohibitively long.  Phase space volume grows exponentially with system size, number of degrees of freedom $L$, and its random exploration time scale would grow exponentially as well.

It is therefore necessary to seek special trajectories that connect the initial and final phase points in as short a time as possible.  The works of Lyapunov~\cite{Lyapunov92} and Poincar\'e~\cite{Poincare99} intimate the solution to this optimization problem.  First, a quantum state provides a coarse grained phase space volume of Planck's constant raised to the power of the number of degrees of freedom, $h^L$~\cite{Wigner32}.  Within such a localized neighborhood of phase space points are initial conditions for trajectories which flee exponentially rapidly from the central phase space point's trajectory.  Likewise, considering the target state's neighborhood, there exist trajectories that are approaching the trajectory associated with the central phase space point exponentially rapidly; i.e., for every positive Lyapunov exponent, there is a negative exponent.  There are invariant sets of phase points associated with the positive and negative Lyapunov exponents called unstable and stable manifolds, respectively.  The intersection of these manifolds defines heteroclinic trajectories.  They have the combined features of fleeing the initial point's trajectory at a maximally exponential rate while simultaneously approaching the final point's trajectory at the maximal exponential rate.  Heteroclinic trajectories exist that can be as close as necessary to the initial and final points.

One of the infinitely many heteroclinic trajectories gives the shortest dynamical connection time between the initial and final phase space points depending on the volume $h^L$, which determines the practical constraint of closeness to initial and final phase space points, although $h^L$ can be chosen as small as preferred.  Denoting this time as $t_{min}$, the immediate neighborhood of trajectories that imitate this heteroclinic motion is of the order of $h^L\exp\left(-h_{KS}t_{min}\right)$, where $h_{KS}$ is Kolmogorov-Sinai entropy~\cite{Kolmogorov58, Sinai59} or for our purposes, the sum of positive Lyapunov exponents. For an extensive system, i.e., the mean positive Lyapunov exponent is constant and $h_{KS}$ grows linearly with the number of degrees of freedom, this neighborhood vanishes exponentially with system size.  Parenthetically, for the mean field Bose-Hubbard Hamiltonian introduced ahead, Eq.~(\ref{eq:clBH_H}), holding the filling factor constant ($N/L$) as the number of sites $L$ increases leads to such an extensive system. Thus, locating a suitable solution leads to a search for an extraordinarily rare trajectory; see Ref.~\cite{Leitao17} for a general technique developed precisely for searching exponentially rare solutions.

Of great interest here is how short the time scale $t_{min}$ can be.  Assuming an extensive system, the available phase space volume per degree of freedom has a geometric mean, ${\cal V}_1$, so that the number of quantum coarse grained cells for the full system is $N_L=({\cal V}_1/h)^L$.  There is a time scale for moving to an adjacent phase space cell, $t_0$, which is also the time for the quantum dynamics to move to the ``closest'' orthogonal state.  The time scale for an ordinary meandering chaotic trajectory to connect two random cells is proportional to $N_Lt_0$, which is an extraordinarily long time.  This is completely different from the optimal heteroclinic trajectory.  It is found somewhere inside a cell stretching as $\exp\left(h_{KS}t\right)$.  This grows to the order of $N_L$ on a very short time scale given approximately by the relation
\begin{equation}
t_{min} \sim \frac{1}{h_{KS}}\ln N_L = \frac{1}{\lambda}\ln \frac{{\cal V}_1}{h} \ ,
\label{eq:tmin}
\end{equation}
where $\lambda = h_{KS}/L$ is the average positive Lyapunov exponent. This is another example of a so-called log-time typically found in chaotic systems.  The quintessential example is the Ehrenfest time at which quantum and classical dynamics must diverge, which is a log-time~\cite{Ehrenfest27, Berman78, Zaslavsky81,Richter22} for a chaotic system.  Depending on the purposes, the details of a particular logtime may vary slightly in terms of Lyapunov exponents (maximum or average, etc..), either classical actions or phase space volumes, but they are all similarly logarithmic in Planck's constant. 

This rather simplistic logic not only gives a vastly shorter time scale than $N_Lt_0$ and one logarithmic in Planck's constant, but surprisingly, independent of the number of degrees of freedom.  Perhaps a more realistic calculation would lead to some weak system size dependence, after all dynamical systems are not uniformly hyperbolic, different cells are more or less difficult to approach, and larger systems occupy larger physical spaces, but clearly even a more realistic estimate would still lead to the conclusion that the optimal heteroclinic motion gives something close to an \textit{exponential} time scale speed-up compared with $N_Lt_0$ and very weak dependence on Planck's constant.  This turns out to be the remarkable way in which chaos becomes a targeting resource.  The cost of this speed up is the difficulty of identifying the requisite \textit{exponentially rare} trajectory.

Without further analysis of how to locate the optimal heteroclinic trajectory, the search space dimension is the immense double-the-degrees-of-freedom of the entire lattice, $2L$, such that any systematic search quickly becomes numerically challenging if not impossible. Conveniently though, chaotic systems usually allow for substantial reduction of the search space, as explicitly demonstrated for a search for heteroclinic trajectories and complex saddles in a Bose-Hubbard system in~\cite{Tomsovic18b}: for any phase space point located inside a chaotic region, the local dynamics can be decomposed into pairs of stable and unstable manifolds. By disregarding all stable directions and those corresponding to constants of the motion, any search in a Bose-Hubbard system can be reduced to an $(L-2)$-dimensional problem. By focusing only on the most unstable directions an even further reduction is usually possible. Even so, a systematic search for general large or higher-dimensional systems is typically futile.

There is a further significant reduction of the search space dimension possible in cases where symmetries are present in the initial and final states; details ahead.  Indeed, this work focuses on Bose-Hubbard systems possessing discrete symmetries, periodic boundary conditions, and symmetric initial conditions leading to a reduced-dimensional symmetry plane of the mean field dynamics~\cite{Steinhuber20}.  The reduced-dimensional dynamics can be mapped onto larger systems with each dimension in the lattice (one, two, or three) being any integer multiple of the reduced-dimensional value, including infinity.  In this way, the search space remains the reduced-dimensional value independent of the size or dimension of the lattice.  Ahead for illustrative purposes, the symmetry reduced-dimensional value of the lattice dimension is $L=4$, and thus an optimized search can be performed in an $L-2=2$-dimensional space no matter what the entire lattice size is.  Thus, it turns out that a simple Monte Carlo search in the reduced phase space around the initial state's centroid is sufficient to determine the suitable heteroclinic pathway,  i.e., control trajectory.  It is assumed in this work that the technical problem of identifying suitable control trajectories is solved and is not discussed further.  Note finally that $t_{min}$ is independent of system size in the case of relying on discrete symmetries, just as Eq.~\eqref{eq:tmin} also gives a system size independent result, i.e., thus the essential motivation of taking advantage of discrete symmetries is the reduced search space.  Furthermore, the control scheme can also be applied to tasks in which a certain time for reaching the target state is given, as long as it is greater than $t_{min}$.  For such larger times,  there  always exists a corresponding hetereoclinic trajectory that can be used as optimal transport path.

\subsection{Bose-Hubbard systems}
\label{sec:bhs}

For many reasons the Bose-Hubbard model turns out to be an ideal bosonic many-body system with which to investigate optimal coherent quantum control.  Not only is this model experimentally relevant for ultracold atoms in an optical lattice, but a classical limit can be defined as a mean-field limit of large occupation numbers and the specific form of the interaction greatly facilitates the control implementation as discussed in Secs.~\ref{sec:bhs} and \ref{subsec:controlHamiltonian} ahead.  A convenient property is that a quadrature phase space representation leads to a Hamiltonian formalism with canonically conjugate real variables.  Furthermore, the control protocol works on a many-body version of minimum uncertainty states in the form of Glauber coherent states, and crucially all the results apply equally well to their total number projected states.  Finally, for the control system, the truncated Wigner approximation becomes a critical analysis tool as its dynamics become essentially exact.

\subsubsection{Hamiltonian and classical (mean field) limit}

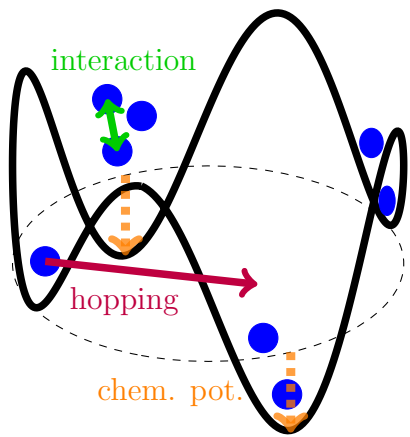
\begin{figure}[t]
	\centering
	\begin{tikzpicture}[tdplot_main_coords,scale=1.3]
	\begin{scope}[color=black, line width=0.1cm,]
	\pgfplothandlerlineto
	\pgfplotfunction{\t}{0,1,...,360}
	{\pgfpointxyz {2*cos(\t)}{2*sin(\t)}{
			cos(4*\t)
			-0.95*exp(-1*((\t-45)*0.001*(\t-45)))
			+0.13*exp(-1*((\t-5)*0.001*(\t-5)))
			-0.95*exp(-1*((\t-225)*0.001*(\t-225)))
	}}
	\pgfusepath{stroke}
	\end{scope}
	\draw [dashed] (2,0,-1) arc[x radius=2, y radius=2, start angle=0, end angle=360];
	\draw[fill,blue] (0.0,0.6,-1.76) circle [x=1cm,y=1cm,radius=0.15];
	\draw[fill,blue] (-0.3,0.75,-2.56) circle [x=1cm,y=1cm,radius=0.15];
	\draw[fill,blue] (.2,1.85,.9) circle [x=0.8cm,y=1cm,radius=0.15];
	\draw[fill,blue] (.2,2.01995,0.25) circle [x=0.55cm,y=1cm,radius=0.15];
	\draw[fill,blue] (.2,-0.915,0.2555) circle [x=1cm,y=1cm,radius=0.15];
	\draw[fill,blue] (.2,-1.025,0.85) circle [x=1cm,y=1cm,radius=0.15];
	\draw[fill,blue] (.2,-0.65,0.725) circle [x=1cm,y=1cm,radius=0.15];
	\draw[<->, black!20!green,line width=0.1cm] (.2,-.915,0.2555)  -- (.2,-1.025,0.85) node[left,yshift= 0.53cm,xshift=1.37506cm ] {interaction};
	\draw[fill,blue] (.2,-1.7,-1.2) circle [x=1cm,y=1cm,radius=0.15];
	\draw[->, purple, line width=0.1cm] (.2,-1.7,-1.2) -- (-0.1,0.5,-1.2)node[midway,below,xshift=-10.2] {hopping};
	\draw[->, orange, line width=0.12cm,opacity=0.75,dashed] (2*0.7071,2*0.7071,-1) -- (2*0.7071,2*0.7071,-1-0.95)node[midway,left,xshift=-.42cm,opacity=1] {chem. pot.};
	\draw[->, orange, line width=0.12cm,opacity=0.75,dashed] (-2*0.7071,-2*0.7071,-1) -- (-2*0.7071,-2*0.7071,-1-0.95)node[midway,left,xshift=-.42cm] {};
	\end{tikzpicture}
	\caption{\textbf{Schematic representation of a $1D$ Bose-Hubbard ring.} Cold atoms trapped in a $1D$ periodic lattice, here consisting of four sites. The model includes nearest-neighbor hopping, on-site interactions between atoms, as well as energetic offsets for each site given by individual chemical potentials.}
	\label{fig:BH_ring}
\end{figure}

A gas of $N$ ultracold atoms subject to a sufficiently strong periodic optical trapping potential generating a lattice may be well described by the Bose-Hubbard model~\cite{Jaksch98}; the lattice can be created in one, two, or three dimensions.  Further ahead, $2D$ lattices are treated, but here the Bose-Hubbard Hamiltonian is only given for a periodic $1D$ lattice of $L$ sites ($L+1 \equiv 1$),
\begin{align}
\begin{split}
\hH = -J \sum_{j=1}^{L} 
\big(
\had{j}\ha{j+1} + \text{h.c.} 
\big)
+ \frac{U}{2}
\sum_{j=1}^{L}
\hat{n}_j \left(\hat{n}_j -1 \right)
+\sum_{j=1}^{L} \mu_j
\hat{n}_j\ .
\end{split}
\label{eq:BH_H}
\end{align}
It energetically accounts for the possible hopping of bosons to neighboring sites, governed by the kinetic hopping parameter $J$, as well as  on-site interactions between the bosons given by an interaction strength $U$, which is related to the s-wave scattering length of the two-body collisions between atoms.  The individual sites can additionally be offset by chemical potentials $\mu_j$; see Fig.~\ref{fig:BH_ring} for a simplified rendition of a $1D$ ring.

As there are two preserved quantities, energy and total particle number, the dimer ($L=2$) is always integrable, as are the limiting cases of $U=0$ or $J=0$, where the absence of an interaction leads to a superfluid ground state and the absence of hopping gives a Mott insulator phase.  For $L > 2$ and in the parameter range $J \sim UN$, the quantum system is known to exhibit quantum chaos~\cite{Kolovsky04,Pausch21}, which is the dynamical regime of current interest.  

A mean field limit, which depends on an effective Planck constant,  can be defined equal to the inverse filling factor, $\hbarE = L/N$.  As $\hbarE \rightarrow 0$, the mean field approximation gets better and better.  A unique Hamiltonian (classical) dynamics arises for the mean field if the ratio 
\begin{equation}
\gamma =  \frac{1}{\hbarE}\frac{U}{J}
\label{eq:gamma} \, , 
\end{equation}
is held fixed.

It is possible to introduce Hermitian operators in terms of $\ha{j}$ and $\had{j}$ known as quadratures that obey position-momentum-like commutation relations,
\begin{equation}
\ha{j} = \frac{(\hat{q}_{j}+i\hat{p}_{j})}{\sqrt{2\hbarE}},\quad \had{j} = \frac{(\hat{q}_{j}-i\hat{p}_{j})}{\sqrt{2\hbarE}}, \quad \left[\hat{q}_j, \hat{p}_j\right] = i\heff\ ,
\end{equation}
which re-expresses the Bose-Hubbard Hamiltonian as
\begin{align}
\begin{split}
\hH = -\frac{J}{\heff} \sum_{j=1}^{L} 
\left(
\hat{q}_j\hat{q}_{j+1} + \hat{p}_j\hat{p}_{j+1} 
\right)
+ \frac{U}{2\heff^2}
\sum_{j=1}^{L}
\left(\frac{\hat{q}_j^2 +\hat{p}_j^2}{2}\right)^2
+\sum_{j=1}^{L} \frac{\mu_j - U}{\heff}
\left(\frac{\hat{q}_j^2 +\hat{p}_j^2}{2}\right) \ .
\end{split}
\label{eq:BH_H2}
\end{align}
Here an irrelevant constant energy offset has been dropped (which is also effectively an $O(\heff^2)$ correction to the spectrum and, in any case, beyond semiclassical argumentation used ahead).  Multiplying both sides of the equation by $\heff/J$ leads to a modified Hamiltonian, 
\begin{align}
\begin{split}
\hH^\prime = - \sum_{j=1}^{L} 
\left(
\hat{q}_j\hat{q}_{j+1} + \hat{p}_j\hat{p}_{j+1} 
\right)
+ \frac{\gamma}{2}
\sum_{j=1}^{L}
\left(\frac{\hat{q}_j^2 +\hat{p}_j^2}{2}\right)^2
+\sum_{j=1}^{L} \mu_j^\prime
\left(\frac{\hat{q}_j^2 +\hat{p}_j^2}{2}\right) \ ,
\end{split}
\label{eq:BH_H3}
\end{align}
where $\mu_j^\prime = (\mu_j - U)/J$.  This Hamiltonian is in a sufficiently symmetric operator form for a full time-dependent WKB approximation~\cite{Maslov81}, i.e., good to $O(\heff^2)$, just through the association of $\hat q \rightarrow q$ and $\hat p \rightarrow p$.  The classical analog Hamiltonian (mean field limit) is given by
\begin{align}
\mathcal{H}\left(\Vec{q}, \Vec{p}\right) = - \sum_{j=1}^{L} 
\left(q_{j}q_{j+1} + p_{j+1}p_{j}\right) + \frac{\gamma}{2}
\sum_{j=1}^{L}\left(\frac{q_{j}^{2} +p_{j}^{2}}{2}\right)^2
+\sum_{j=1}^{L} \mu_j^\prime\left(\frac{q_{j}^{2} +p_{j}^{2}}{2}\right)\ ,
\label{eq:clBH_H}
\end{align}
where the degree of chaos is determined by the value of $\gamma$.  Typically, the strongest chaos is approximately in the range $1 < \gamma < 3$; see~\cite{Kolovsky04, Tomsovic18b}.  The dynamics of the quadratures is thus governed by Hamilton's equation of motion 
\begin{align}\label{eq:Ham_equations}
\dot{\Vec{x}} =  \mathbb{\Sigma} \,\vec{\nabla}_{\Vec{x}} \mathcal{H}\ , \quad \text{ with } \Vec{x}= (\Vec{q},\Vec{p}) \text{ and } \mathbb{\Sigma} = \begin{pmatrix} 0 &\mathbb{1} \\ -\mathbb{1} & 0   \end{pmatrix}\ ,
\end{align}
in a 2$L$-dimensional phase space.
The classical dynamics conserves energy and total number of particles along each trajectory, i.e., each mean-field solution. For any targeting application using the Bose-Hubbard model this implies that the space of possible targeting states is restricted to those associated to phase space points sharing the same energy and particle number surface as the corresponding initial state.  Choosing the initial chemical potentials  $\mu_j^\prime$ of the individual sites however allows for quite a bit of freedom in putting initial and target states on the same energy surface, so the constraint posed by energy conservation can be largely evaded.

\subsubsection{Localized states}
\label{sec:locstat}

Similar to~\cite{Tomsovic23,Tomsovic23b}, the protocol for Bose-Hubbard systems is designed to control states that are initially localized in the quadrature phase space.  Glauber coherent states provide a natural and optimal choice with which to embark as they are minimum uncertainty states and thus the most classical possible.  Defined as eigenstates of the annihilation operator
\begin{equation}\label{eq:coherentStateEigenstate}
\hat{\Vec{a}}\coh{\vphi} = {\vphi}\coh{\vphi}\ ,
\end{equation}
they can be written explicitly as
\begin{align}
\coh{\vphi} = \exp{-\frac{\lVert{{\vphi}}\rVert^{2}}{2} + \vphi \cdot \vad} \ket{0}\ .
\label{eq:CohState}
\end{align}
Their phase space representation is given by the Wigner function and is well known to be a simple symmetric $2L$-dimensional Gaussian wave packet 
\begin{equation}
\label{eq:wignerDist}
\mathcal{W}\left(\Vec{q}, \Vec{p}\right) = \left(\pi\hbar_{\text{eff}}\right)^{-L}\exp{-\frac{\left(\Vec{q}-\Vec{q}_{\alpha}\right)^2}{\hbar_{\text{eff}}}-\frac{\left(\Vec{p}-\Vec{p}_{\alpha}\right)^2}{\hbar_{\text{eff}}}}\ , 
\end{equation}
centered around the mean-field point $\Vec{\phi}_{\alpha} = (\Vec{q}_{\alpha} + i\Vec{p}_{\alpha})/\sqrt{2\heff}$.
The coherent state variance per degree of freedom is given by $\sigma^2 = \heff/2$, which for a single degree of freedom has the $2\sigma$-contour enclosing a volume $2\pi\heff$.  More generally, the volume per quantum state is  $(2\pi\heff)^L$~\cite{Wigner32}, as mentioned above.  

The Bose-Hubbard system possesses a $U(1)$ dynamical symmetry, which in mean field quadratures equates to an $SO(2)$ symmetry.  Each trajectory maps perfectly onto another by simultaneously rotating each pair $(q_j,p_j)$ through any given angle $0\le \theta < 2\pi$. This symmetry has the consequence that it implies a generalization of the control protocol in a straightforward manner. 

The same scheme as for a Glauber coherent state also works for a number projected coherent state
\begin{align}
\proj{\vphi} = \frac{1}{\sqrt{\lVert{{\vphi}}\rVert^{2N} N!}} \big(\vphi \cdot \vad\big)^{N} \ket{0}\ ,
\label{eq:ProjState}
\end{align}
which contains a fixed number of particles, and is a Fock state in some basis; states of this kind can be constructed~\cite{Greiner02b}.  This additional class of states is of great physical significance just by virture of fixing the particle number, but it also contains interesting cases such as $\phi^{\alpha}_{j} = \exp\left(2\pi i\alpha j/L\right)$ with $\alpha=0$ corresponding to the non-interacting ground state of the superfluid phase and $\alpha=1$ being the first excited state of the momentum operator. Another class of states that can be represented as number projected coherent states are all Fock states concentrated on a single site $k$, e.g., $\phi_{j} = \sqrt{L}\delta_{j,k}$.\\


\subsubsection{Truncated Wigner Approximation}
\label{sec:twa}

Due to their simpler representation, the discussion follows coherent states, however all results are equally applicable to the number projected counterparts. In quadratures, the propagation of the state is given by the Moyal bracket~\cite{Moyal49}
\begin{align}
\label{eq:moyal}
\begin{split}
\frac{d}{dt} \mathcal{W}\left(\Vec{q}, \Vec{p}\right) &= \left\{ \mathcal{H}\left(\Vec{q}, \Vec{p}\right), \mathcal{W}\left(\Vec{q}, \Vec{p}\right) \right\}_{\text{MB}}= \frac{2}{\heff}\mathcal{H}\left(\Vec{q}, \Vec{p}\right)\sin\left[\frac{\heff\overleftrightarrow{\Lambda}}{2}\right] \mathcal{W}\left(\Vec{q}, \Vec{p}\right)\ ,
\end{split}
\end{align}
with $\overleftrightarrow{\Lambda} = \frac{\overleftarrow{\partial}}{\partial\Vec{q}}\frac{\overrightarrow{\partial}}{\partial\Vec{p}} - \frac{\overleftarrow{\partial}}{\partial\Vec{p}}\frac{\overrightarrow{\partial}}{\partial\Vec{q}}$ denoting the symplectic operator, which acts like the Poisson bracket. An important consequence of Eq.~\eqref{eq:moyal} is that for any quadratic Hamiltonian all higher order terms vanish, which dovetails perfectly with the control protocol introduced ahead.

Dropping third and higher order terms in $\heff$ yields the truncated Wigner approximation (TWA)~\cite{Steel98, Sinatra02} in which quantum zero-point fluctuations are encoded in the Wigner transforms, but otherwise quantum interference between mean field paths is dropped. The distribution of initial conditions then evolves under a classical Liouville equation 
\begin{equation}
\frac{d}{dt} \mathcal{W}\left(\Vec{q}, \Vec{p}\right) = \left\{ \mathcal{H}\left(\Vec{q}, \Vec{p}\right), \mathcal{W}\left(\Vec{q}, \Vec{p}\right) \right\}\ .
\end{equation}
The Gaussian form for coherent states of Eq.~(\ref{eq:wignerDist}) is straightforward to implement within the TWA.  This renders the TWA an excellent analysis tool for this work. 

\subsubsection{Density Wave}
\label{sec:dw}

\begin{figure}
	\centering
	\begin{subfigure}{0.49\textwidth}
		\centering
		\includegraphics[width = \textwidth]{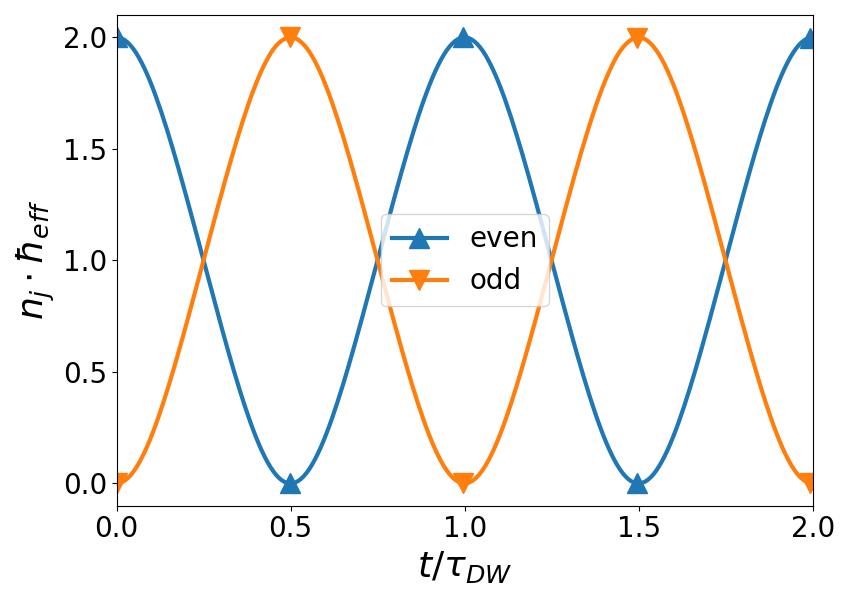}
		\caption{}
		\label{fig:periodsDWa}
	\end{subfigure}
	\begin{subfigure}{0.49\textwidth}
		\centering
		\includegraphics[width = \textwidth]{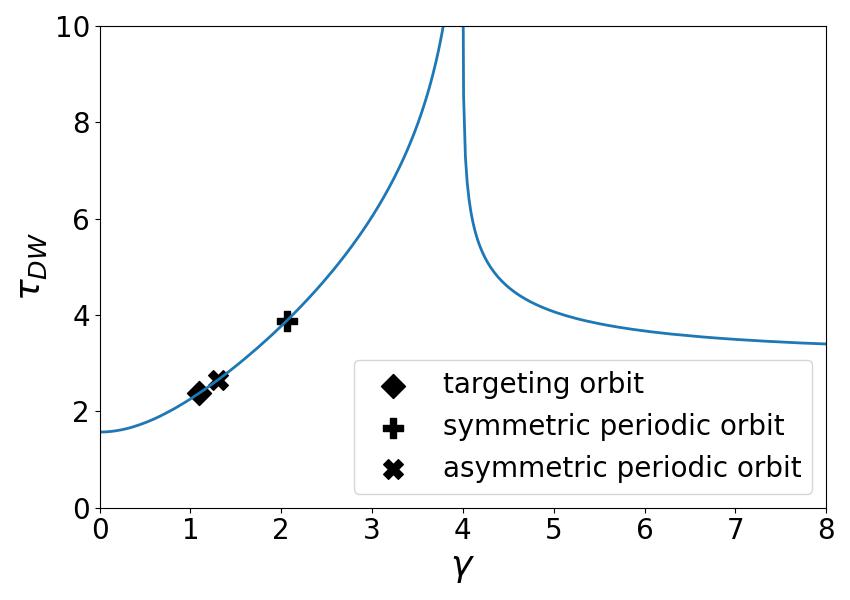}
		\caption{}
		\label{fig:periodsDW}
	\end{subfigure}
	\caption{\textbf{Properties of a density wave.} A density wave initially centered around $\Vec{\phi} = \sqrt{2/\heff}\left(1,0,1,0\right)$ exhibits partial or full periodic population inversion with a period $\tau_{DW}$ as illustrated in \textbf{(a)}.  How the period depends on dynamical parameter $\gamma$, Eq.~(\ref{eq:gamma}), of the system is shown in \textbf{(b)}. Except for $\gamma = 4$ where the period diverges since $\Vec{\phi}$ lies on a separatrix of the mean field dimer, $\tau_{DW}$ provides a well defined time scale of the system. The time scale is marked for all the specific trajectories used throughout the work. }
	\label{fig:DensityWave}
\end{figure}

Density waves corresponding to coherent states centered on mean-fields of the form $\Vec{\phi} = \sqrt{2/\heff}\left(1,0,1,0...\right)$, alternating sites for higher-dimensional lattices,  are special solutions to Hamilton's equations given by initially populating every second site equally, while leaving the rest unoccupied.  The mean field solutions, using initial conditions based on their centroid values, can be mapped to and from the dimer~\cite{Steinhuber20}, which is an integrable system. The resulting trajectories exhibit periodic population inversion for $\gamma<4$ with a period $\tau_{DW}(\gamma)$, see Fig.~\ref{fig:periodsDWa}. For $\gamma>4$, the system is in a self-trapping regime where the solution is still periodic, but only a partial population imbalance is achieved.  If $\Vec{\phi}$ lies on the symmetry plane separatrix of the dimer, $\gamma=4$, the period diverges as shown in Fig.~\ref{fig:DensityWave}. 
Due to these special properties, it serves as a convenient initial state for the applications discussed throughout this work, even though theoretically any other state with a centroid located in a chaotic phase space region could be used. This makes $\tau_{DW}$ a natural choice for a dynamical time scale, as long as the systems considered are sufficiently removed from the singularity at $\gamma = 4$, as is done ahead.  The corresponding time scales $\tau_{DW}$ of this work's specific cases are depicted in Fig.~\ref{fig:periodsDW}.

\subsection{Control Hamiltonian}
\label{subsec:controlHamiltonian}

Given that a (presumably heteroclinic) control trajectory $\left(\Vec{q}(t), \Vec{p}(t)\right)_c$ connecting the mean-field centers of the initial and target state at time $t_{min}$, respectively, is identified, the initial goal is to guide the coherent state along this solution. \begin{figure}
	\centering
	\begin{subfigure}{\textwidth}
		\centering
		\hspace{-0.7cm}$t=0$
		\includegraphics[width = \textwidth]{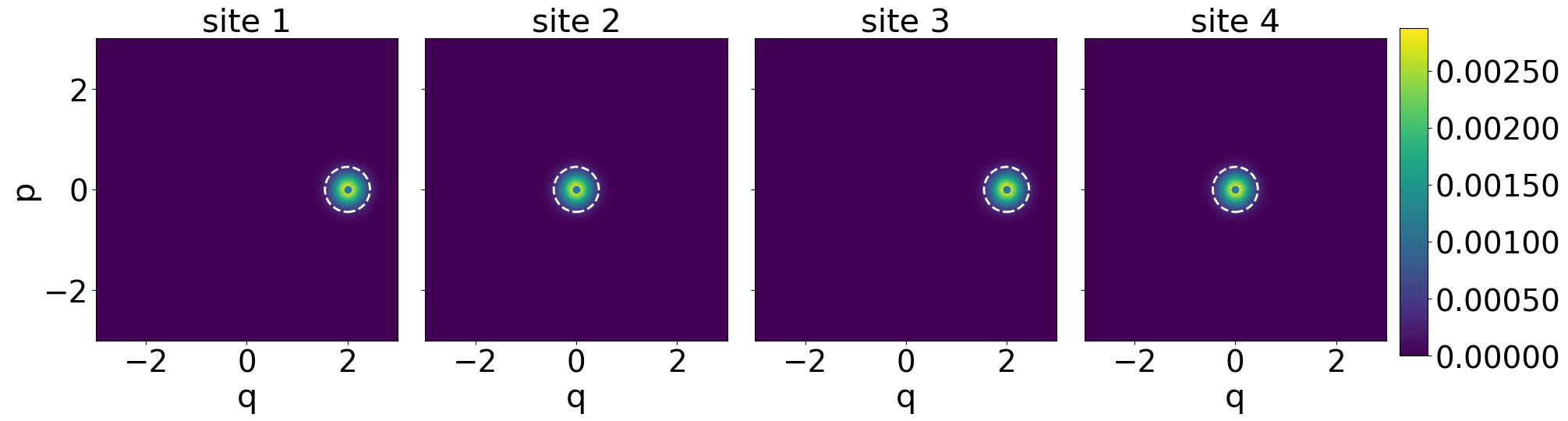}
	\end{subfigure}
	\begin{subfigure}{\textwidth}
		\centering
		\hspace{0.3cm}$ t=0.5\tau_{\text{DW}}$
		\includegraphics[width = \textwidth]{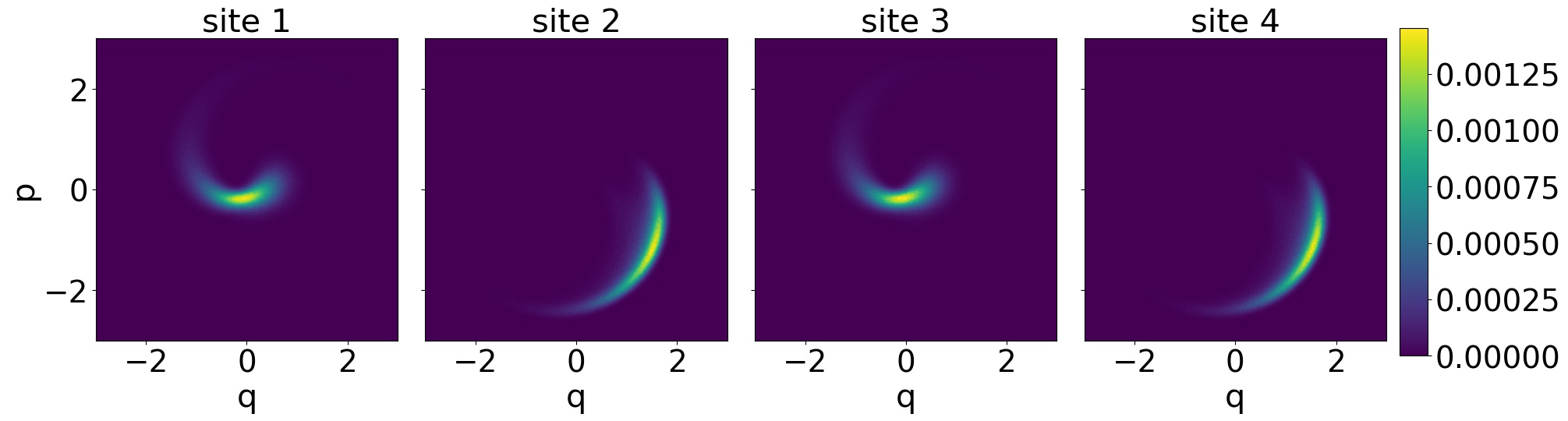}
	\end{subfigure}
	\begin{subfigure}{\textwidth}
		\centering
		\hspace{-0.2cm}$t=\tau_{\text{DW}}$
		\includegraphics[width = \textwidth]{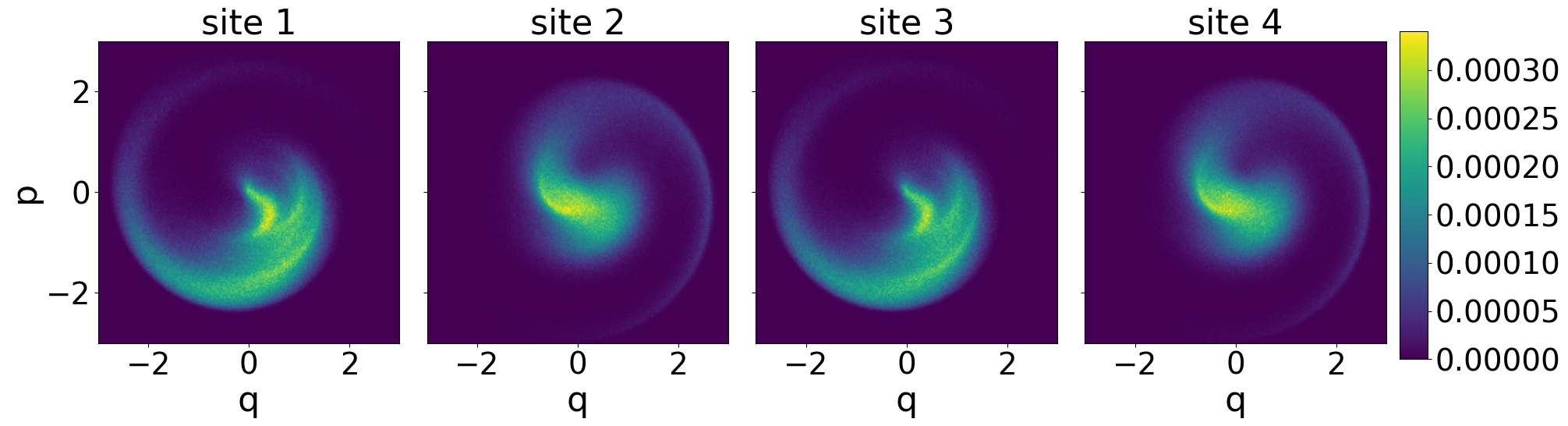}
	\end{subfigure}
	\begin{subfigure}{\textwidth}
		\centering
		$t=4\tau_{\text{DW}}$
		\includegraphics[width = \textwidth]{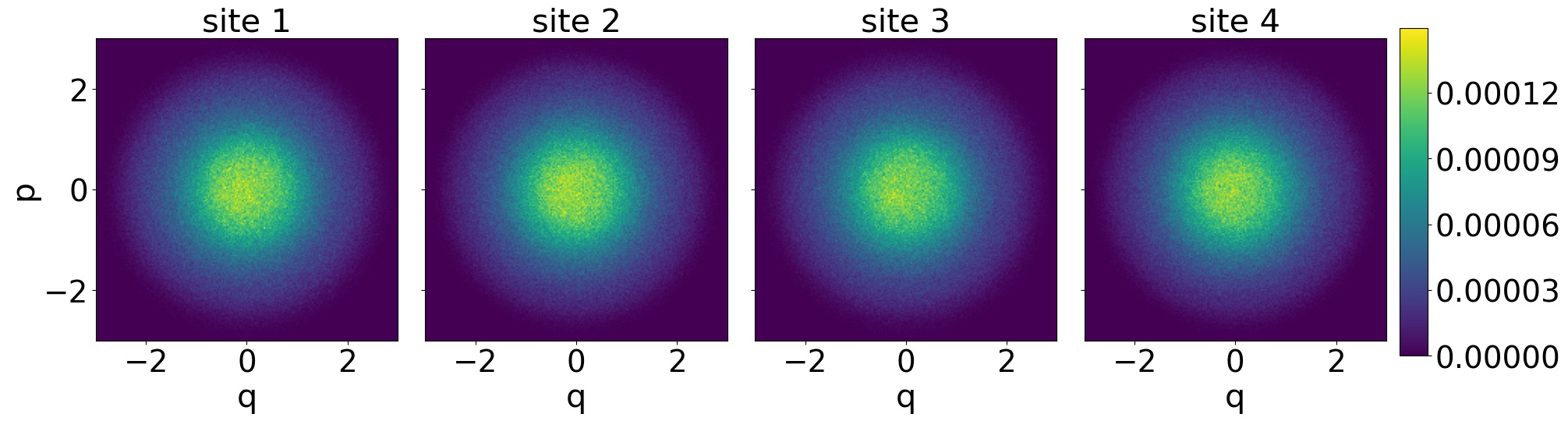}
	\end{subfigure}
	\caption{\textbf{Spreading of a localized coherent state in a Bose-Hubbard system.}  A minimum uncertainty coherent state (here for illustration chosen to be $\heff = 1/10$), initially centered on the density wave $\Vec{\phi}_{\alpha} = \sqrt{2/\heff}\left(1,0,1,0\right)$ (indicated as dot in the upper panel) is propagated in the Bose-Hubbard system 
		($\gamma \simeq 1.965$) using the TWA. With the scaled dynamics that is used, the mean-field corresponds to the phase space point $\Vec{q}_{\alpha}=(2,0,2,0)$ and $\Vec{p}_{\alpha} = (0,0,0,0)$, independent of $\heff$. 
		The state, whose volume is for $t = 0$ well contained in the $2\sigma$-contour (white dashed circle), quickly spreads, already being far from its initial Gaussian shape after one period of the density wave. At $t = 4\tau_{DW}$ the state is fully equilibrated.}
	\label{fig:BH_equilibration}
\end{figure}
Due to the quartic polynomial in quadratures present in the interaction term of Eq.~(\ref{eq:clBH_H}), the initially localized state spreads exponentially in time as the dynamics are strongly chaotic in the dynamical regimes of interest.  This spreading is so rapid that without some control mechanism, a quantum targeting is impossible beyond an exceedingly short logarithmic time scale in $\heff$.  Figure~\ref{fig:BH_equilibration} demonstrates this fast equilibration for the example of a coherent state initially centered on a density wave.  The full time evolution is provided in Animation 1 in the supplementary material.

It is theoretically possible to proceed as in~\cite{Tomsovic23} and apply an additional unitary transformation to the system sufficiently often that is designed to counteract the spreading.  However, this generally introduces terms in the time evolution that are not particle number conserving and thus challenging, if not impossible, with respect to experimental realization. There exists a much simpler approach developed in~\cite{Tomsovic23b} based on introducing a new time-dependent simulation control Hamiltonian, $\hat{H}_c$, that is uniquely designed for some particular targeting problem with fixed initial and final quadrature conditions; note that a minimal time related to heteroclinic motion could be the desired goal or any specific time longer than that.  

Consider Hamilton's equations for each component using Eq.~\eqref{eq:clBH_H},
\begin{align}
\begin{split}
&\dot{q}_j = \frac{\partial{\mathcal{H}}\left( \Vec{q}, \Vec{p} \right) }{\partial{p_j}}\biggr\rvert_{\Vec{q}(t), \Vec{p}(t)} = \left[\mu_j^\prime + \gamma \left(\frac{q_j^2+p_j^2}{2}\right) \right] p_j - p_{j+1} - p_{j-1} \, ,  \\
&\dot{p}_j = -\frac{\partial\mathcal{H}\left(\Vec{q}, \Vec{p}\right)}{\partial{q_j}}\biggr\rvert_{\Vec{q}(t), \Vec{p}(t)} = -\left[\mu_j^\prime + \gamma \left(\frac{q_j^2+p_j^2}{2}\right) \right] q_j + q_{j+1} + q_{j-1} \ .
\label{eq:conBH_eqM}
\end{split}
\end{align}
Evaluated using the control trajectory, $\left(\Vec{q}(t), \Vec{p}(t)\right)_c$, the terms in large square parentheses can be replaced by time-dependent values that are no longer functions of $\{q_j,p_j\}$.  With that replacement, it is straightforward to see that a control Hamiltonian can be created with the interaction term switched off given by
\begin{equation}
\mathcal{H}_c\left(\Vec{q}, \Vec{p}, t\right) = - \sum_{j=1}^{L} 
\left(q_{j}q_{j+1} + p_{j+1}p_{j}\right) +\sum_{j=1}^{L} \mu_j(t) \left(\frac{q_{j}^{2} +p_{j}^{2}}{2}\right) \ ,
\label{eq:clBH_Hcont}
\end{equation}
which is in essence a harmonic oscillator with chemical potentials that are time-dependent functions of the control trajectory's site occupancies,
\begin{equation}
\label{eq:tdcp}
\mu_j(t) = \mu_j^\prime + \gamma \left(\frac{q_{j}^{2}(t) +p_{j}^{2}(t)}{2}\right) = \mu_j^\prime + \gamma n_j(t) \ .
\end{equation}
The control Hamiltonian makes use of the chemical potentials $\mu_j(t)$ as purely time-dependent functions evaluated along the chosen control (heteroclinic) trajectory $(\vec{q}(t),\vec{p}(t))_c$, which effectively replaces the originally cubic terms in quadratures with time-dependent linear terms.  In this way, the control trajectory and only the control trajectory, $(\vec{q}(t),\vec{p}(t))_c$ is a solution of Hamilton's equations for both Hamiltonians $\mathcal{H}\left(\Vec{q}, \Vec{p}\right)$ and $\mathcal{H}_c\left(\Vec{q}, \Vec{p}, t\right)$.  In addition, $\mathcal{H}_c\left(\Vec{q}, \Vec{p}, t\right)$ having a quadratic form leads to the TWA being exact for its quantized version,
\begin{align}
\begin{split}
\hH_{\text{c}}(t) = - \sum_{j=1}^{L}& 
\left(\had{j}\ha{j+1} + \had{j+1}\ha{j}
- \mu_{j}(t)\hat{n}_j \right) \ ,
\end{split}
\label{eq:conBH_H}
\end{align}
and there is no longer any spreading in the neighborhood of $(\vec{q}(t),\vec{p}(t))_c$, or for that matter, anywhere else.  Although, $(\vec{q}(t),\vec{p}(t))_c$ is an exponentially unstable solution of $\mathcal{H}\left(\Vec{q}, \Vec{p}\right)$, it is a stable solution of $\mathcal{H}_c\left(\Vec{q}, \Vec{p}, t\right)$.

Up to this point there is a broad freedom in the choice of a particular chaotic Hamiltonian that may serve as a black box to produce a desired control solution.  The apparently natural choice of a Bose-Hubbard type of system, however, points to a deeper connection between the dispersionless dynamics generated by Eq.~(\ref{eq:conBH_H}) and the precise form of the interaction (quartic) term in Eq.~(\ref{eq:clBH_H}).  In addition, this choice is quite favorable from the point of view of a practical implementation, since among the infinity of nonlinear systems that one might consider for producing a control trajectory, a key selection criteria is the desire to have a minimal alteration of a realistic experimental set up. The Bose-Hubbard Hamiltionian, describing a broad range of self-interacting bosonic systems, satisfies this condition thanks to the technical possibilities to tune the interaction~\cite{Chin10, Su23, Impertro23} and chemical potential parameters, and to implement Eq~(\ref{eq:conBH_H}). 

There is however another deep conceptual reason for the special connection between the control and Bose-Hubbard Hamiltonians. It relies on the coherent representation of the time evolution operator for Bose-Hubbard systems
\begin{equation}
K(\vphi_{\beta},\vphi_{\alpha};t)=\langle   \vphi_{\beta} \left| {\rm e}^{-\frac{i}{\hbar}\hat{H}t} \right|\vphi_{\alpha}\rangle,
\end{equation}
with $\hat{H}$ given in Eq.~(\ref{eq:BH_H}), in terms of auxiliary fields
\begin{equation}
\label{eq:fin_text}
K^{\rm sc}(\vphi_{\beta},\vphi_{\alpha};t)=\sum_{\gamma} \frac{{\cal D}_{\gamma}(\vphi_{\alpha},\vphi_{\beta},t)}{Z({\bf v})} {\rm e}^{\frac{i}{2}\int_{0}^{t}ds\vec{\sigma}_{\gamma}(s){\bf v}^{-1}\vec{\sigma}_{\gamma}(s)} \times K_{\vec{\sigma}_{\gamma}} (\vphi_{\beta},\vphi_{\alpha};t) 
\end{equation}
valid in the limit of large mean occupations $N\gg 1$. Here $K_{\vec{\sigma}_{\gamma}} (\vphi_{\beta},\vphi_{\alpha};t)$ is the transition amplitude between initial $|\vphi_{\alpha}\rangle$ and target $|\vphi_{\alpha}\rangle$ coherent states defined by the control Hamiltonian, Eq.~(\ref{eq:conBH_H}), with control field $\vec{\mu}(t)=\vec{\sigma}_{\gamma}(t)$. As shown in~\ref{sec:A1}, the total transition amplitude for the full interacting problem is given as a coherent sum, weighted by the factors $\frac{{\cal D}_{\gamma}(\vphi_{\alpha},\vphi_{\beta},t)}{Z({\bf v})} {\rm e}^{\frac{i}{2}\int_{0}^{t}ds\vec{\sigma}_{\gamma}(s){\bf v}^{-1}\vec{\sigma}_{\gamma}(s)}$, over the set of control fields $\vec{\sigma}_{\gamma}(s)$ obtained by $\sigma_{\gamma,j}(s)=\phi_{\gamma,j}(s)\phi^{*}_{\gamma,j}(s)$ from the solutions of the mean field boundary problem $\vphi_{\gamma}(s)$. This representation leads to a natural interpretation of the control Hamiltonian as a contribution to the full interacting problem, that is selected based on further physical considerations. One such physical consideration, relevant for the approach discussed here, is the real character of the control fields. Since, generically, the solutions of the mean field problem linking initial and target coherent states admit only solutions with complexified quadratures and the corresponding complexified auxiliarly fields, one instead looks for nearby states admiting one solution with real quadratures, making the control Hamiltonian Hermitian. 

Furthermore, by its very construction, the transition amplitude for a given auxiliary field is dispersionless due to the quadratic form of the control Hamiltonian. Specifically, this linearity of the control evolution $\hH_{\text{c}}(t)$, implies that the coherent state $\coh{\vphi(t)}$ (possibly number projected) centered on the mean-field solution used to drive the system satisfies the time-dependent Schrödinger equation, i.e. 
\begin{align}
i\frac{\partial}{\partial t} \coh{\vphi(t)} =  \hH_{\text{c}}(t)  \coh{\vphi(t)}\ .
\end{align}
The explicit proof is provided in \ref{sec:A2}.
Given the simple form of Eq.~(\ref{eq:conBH_H}), the actual realization of the protocol requires switching off the interactions \cite{Chin10, Su23, Impertro23} and controlling the chemical potentials of each individual site in a time-dependent fashion. Some aspects of the protocol robustness to imperfections are addressed in Sec.~\ref{subsec:errors}.

\section{Coherent quantum targeting in 1D lattices}
\label{sec:1d_applications}


Here, a few representative example applications of the protocol on a 1D periodic lattice are presented.  Preparing states with well-defined phase relations between the condensates on different sites is covered in Sec.~\ref{subsec:targeting} and the creation of non-trivial periodic evolution in Sec.~\ref{subsec:periodic}.  As it turns out, even though the idea of placing quantum states on classical trajectories is semiclassical in nature, the control system is harmonic and thus behaves purely classical. This implies that the method should work perfectly well for both the large particle number limit, as well as in dilute quantum regime. However, in this treatment the small shifts between initial and target quadrature centers to the initial and final points of the control trajectory, $(\vec{q}(t),\vec{p}(t))_c$, are not taken into account.  The resulting effects as well as imperfections in making the residual interactions vanish both introduce an $\heff$ dependency to the fidelity of the protocol and are addressed in Sec.~\ref{subsec:errors}.  Both the TWA and full quantum simulations of the time evolution are given. 

\subsection{Targeting}
\label{subsec:targeting}

\begin{figure}
	\centering
	\begin{subfigure}{0.49\textwidth}
		\centering
		\begin{tikzpicture}[tdplot_main_coords,scale=1.]
		\begin{scope}[color=black, line width=0.1cm,]
		\pgfplothandlerlineto
		\pgfplotfunction{\t}{0,1,...,360}
		{\pgfpointxyz {2*cos(\t)}{2*sin(\t)}{
				cos(4*\t)
				-0.95*exp(-1*((\t-45)*0.001*(\t-45)))
				+0.13*exp(-1*((\t-5)*0.001*(\t-5)))
				-0.95*exp(-1*((\t-225)*0.001*(\t-225)))
		}}
		\pgfusepath{stroke}
		\end{scope}
		\draw [dashed] (2,0,-1) arc[x radius=2, y radius=2, start angle=0, end angle=360];
		\draw[fill,blue] (0.0,0.6,-1.76) circle [x=1cm,y=1cm,radius=0.15];
		\draw[fill,blue] (-0.3,0.75,-2.56) circle [x=1cm,y=1cm,radius=0.15];
		
		\draw[fill,blue] (.2,-0.815,-0.5555) circle [x=1cm,y=1cm,radius=0.15];
		\draw[fill,blue] (.2,-0.65,0.235) circle [x=1cm,y=1cm,radius=0.15];
		
		\end{tikzpicture}
		\caption{}
	\end{subfigure}
	\begin{subfigure}{0.49\textwidth}
		\hspace{1.4cm}
		\begin{tikzpicture}[tdplot_main_coords,scale=1.]
		\begin{scope}[color=black, line width=0.1cm,]
		\pgfplothandlerlineto
		\pgfplotfunction{\t}{0,1,...,360}
		{\pgfpointxyz {2*cos(\t)}{2*sin(\t)}{
				cos(4*\t)
				-0.95*exp(-1*((\t-45)*0.001*(\t-45)))
				+0.13*exp(-1*((\t-5)*0.001*(\t-5)))
				-0.95*exp(-1*((\t-225)*0.001*(\t-225)))
		}}
		\pgfusepath{stroke}
		\end{scope}
		\draw [dashed] (2,0,-1) arc[x radius=2, y radius=2, start angle=0, end angle=360];
		\draw[fill,blue] (-0.3,0.75,-2.56) circle [x=1cm,y=1cm,radius=0.15];
		
		\draw[fill,blue] (.2,2.01995,0.35) circle [x=0.55cm,y=1cm,radius=0.15];
		\draw[fill,blue] (.2,-0.815,-0.5555) circle [x=1cm,y=1cm,radius=0.15];
		
		\draw[fill,blue] (.2,-1.7,-1.2) circle [x=1cm,y=1cm,radius=0.15];
		\draw [->,orange,ultra thick] (-0.5,-.71,0.32) to [out=30,in=150] node[midway,below]{${\pi/2}$} (-1.,1.4,0.32) ;
		\draw [->,orange,ultra thick] (-1.,2.0,-0.3)  to [out=-20,in=25] node[midway,below,right]{${\pi/2}$} (2.,2.2,-1.02) ;
		\draw [->,orange,ultra thick] (2.,0.95022,-1.402) to [out=190,in=260] node[midway,below]{${\pi/2}$} (2.,-1.32,-0.82);
		\draw [->,orange,ultra thick] (2.,-1.62,-0.42) to [out=150,in=160] node[midway,left]{${\pi/2}$} (.2,-1.16,0.12) ;
		\end{tikzpicture}
		\caption{}
	\end{subfigure}
	\caption{\textbf{Schematic of initial and target quantum state.} Raising the unoccupied sites energetically by choosing 
		$\Vec{\mu} = (0, \gamma, 0, \gamma)$ puts the density wave, (a), and excited momentum eigenstate (as target), (b), on the same classical energy surface. This opens up the possibility of finding a control trajectory, connecting both mean fields in quadrature phase space. The corresponding initial coherent quantum state can be guided along the control trajectory into the target quantum state by driving the control system with the associated time-dependent occupation numbers. }
	\label{fig:orbitTargeting1}
	
\end{figure}
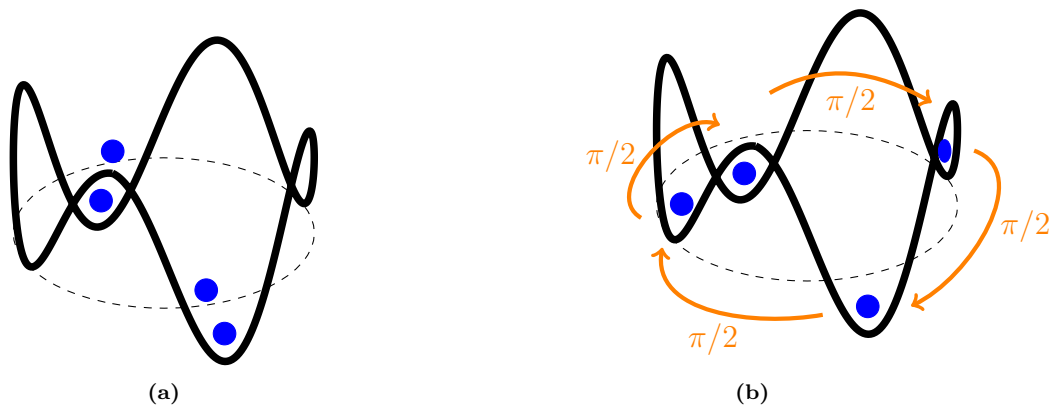

As discussed in Sec.~\ref{sec:chaos}, the great advantage of using heteroclinic pathways of the chaotic mean-field limit as guiding control trajectories is their exponential speed-up of connecting any two points in phase space provided by ergodicity.  In principle, any state of the form of Eqs.~\eqref{eq:CohState} or \eqref{eq:ProjState} can be prepared by targeting the corresponding mean-field starting from an initial field sharing the same energy and particle number surfaces. This includes states where the condensates on each site carry well defined, non-trivial phases~\cite{Impertro23}.   As a proof of principle, the targeted many-body state is chosen to have the bosons occupying the first excited momentum eigenstate on a four-site ring, which corresponds to the homogeneous mean-field $\Vec{\phi}_{\beta} = \sqrt{1/\heff}(1,i,-1,-i)$ with $\pi/2$-phases between neighboring sites.

\begin{figure}
	\centering
	\includegraphics[width = \textwidth]{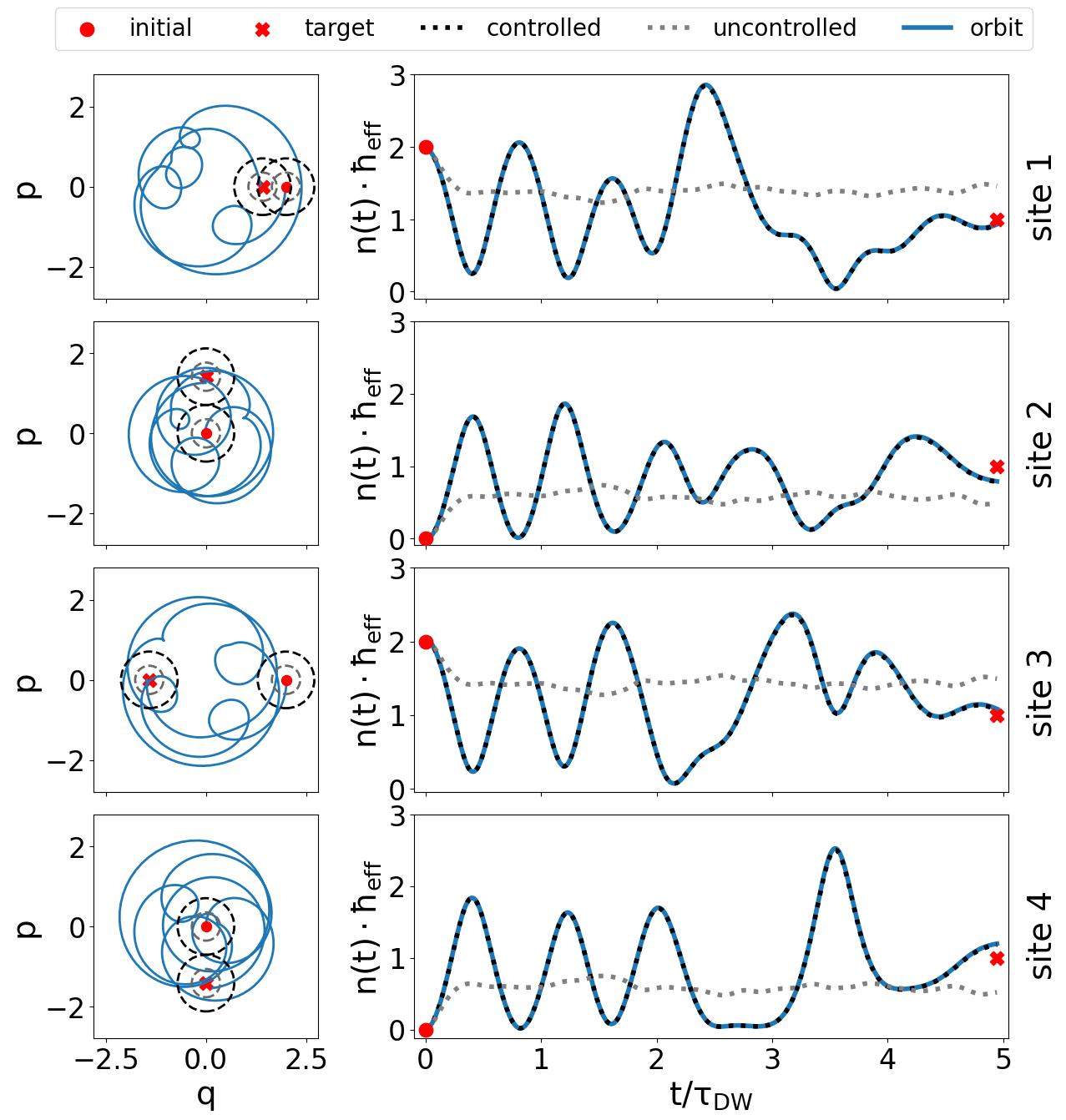}
	\caption{\textbf{Control Trajectory to target excited momentum eigenstate.} The trajectory connecting the centroids of the initial density wave $\Vec{\phi}_{\alpha} = \sqrt{2/\heff}(1,0,1,0)$ (red dot) with the final target state $\Vec{\phi}_{\beta} = \sqrt{1/\heff}(1,i,-1,-i)$ (red cross) is shown for the different sites in the left panel ($\gamma \simeq 1.965$).  The respective $1\sigma$-contours for the initial and target coherent state's Wigner densities, Eq.~\eqref{eq:wignerDist}, are depicted centered on their respective quadrature centroids with $\heff=1$ (black  dashed circles ) and $\heff=1/4$ (gray  dashed circles). Generally, the area inside the circles shrinks proportionally to $\heff$. The time-dependent control chemical potentials shown in the right panel are determined by the respective site occupations by Eq.~\eqref{eq:tdcp} and the offsets of $\mu_j^\prime$ given in the text in Sec.~\ref{subsec:targeting}.  Initially, somewhat similar to the integrable behavior of the density wave, the chaotic nature of the control trajectory quickly dominates and it reaches the target at $t = 4.9\tau_{DW}$. Additionally, the expectation values of the occupations are shown for the quantum propagation of the coherent state evolving under the original Bose-Hubbard Hamiltonian (uncontrolled, grey dotted), which deviate rapidly from the classical solution.  Under the control Hamiltonian (controlled, black dotted) the expectation values shadow the control trajectory, as they must.}
	\label{fig:orbitTargeting}
\end{figure}

As already discussed, the protocol is initialized with a coherent state centered on a density wave $\Vec{\phi}_{\alpha} = \sqrt{2/\heff}(1,0,1,0)$. By choosing the chemical potential offset accordingly, i.e. $\Vec{\mu}^\prime = (0, \gamma, 0, \gamma)$, both mean-fields share the same classical energy surface for every value of $\gamma$ and can be connected with a transport trajectory, see Fig.~\ref{fig:orbitTargeting1}.  Figure~\ref{fig:orbitTargeting} shows a heteroclinic trajectory with a control time of less then five oscillations of the density wave $\tau_{DW}$ (Fig.~\ref{fig:periodsDW}), which is quite short.  If greater precision is required, then ergodicity and the nature of unstable and stable manifolds guarantees the existence of heteroclinic trajectories closer to the initial and target mean-fields at the expense of a longer control time.
The initial state $\coh{\Vec{\phi}_{\alpha}}$ then evolves under the control Hamiltonian, Eq.~(\ref{eq:conBH_H}), by driving the system with the time dependent chemical potentials that are determined by the respective site occupations shown in the right panel of  Fig.~\ref{fig:orbitTargeting}.  The time-dependent site chemical potentials follow using Eq.~\eqref{eq:tdcp} and the definition of $\mu_j^\prime$ given in the text immediately after Eq.~\eqref{eq:BH_H3}. \change{A TWA simulation (quantum calculation) is used to calculate the propagation of the state.  Note that the TWA calculation exactly matches the quantum calculation due to the quadratic nature of the control Hamiltonian, and they cannot be distinguished in this case.}

\begin{figure}
	\centering
	\begin{subfigure}{\textwidth}
		\centering
		$t = 0$
		\includegraphics[width = \textwidth]{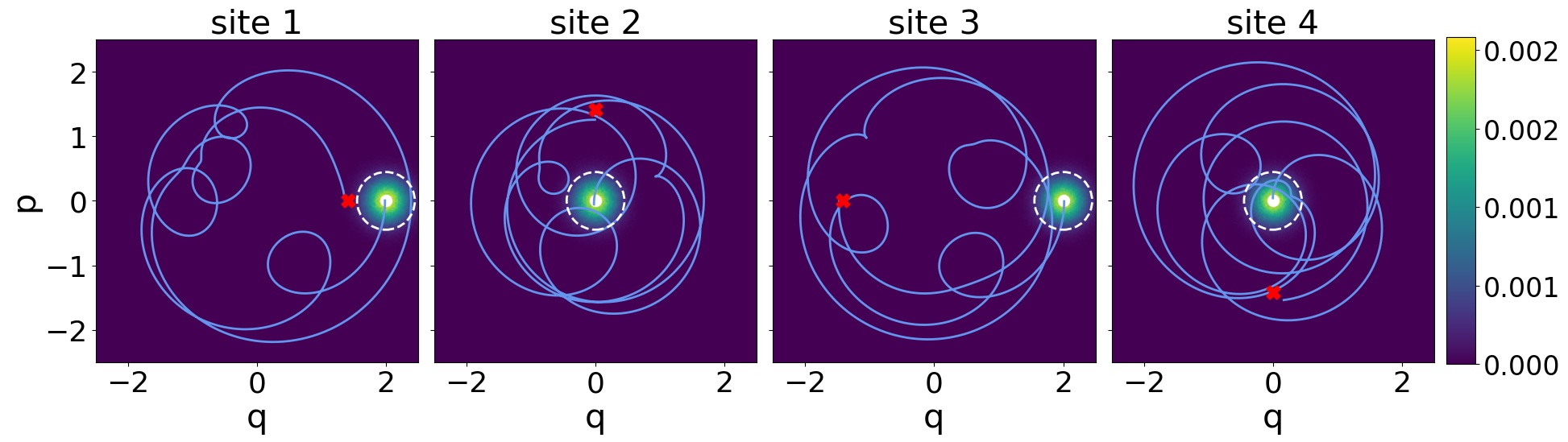}
	\end{subfigure}
	\begin{subfigure}{\textwidth}
		\centering
		$t =4.9\tau_{\text{DW}}$
		\includegraphics[width =    \textwidth]{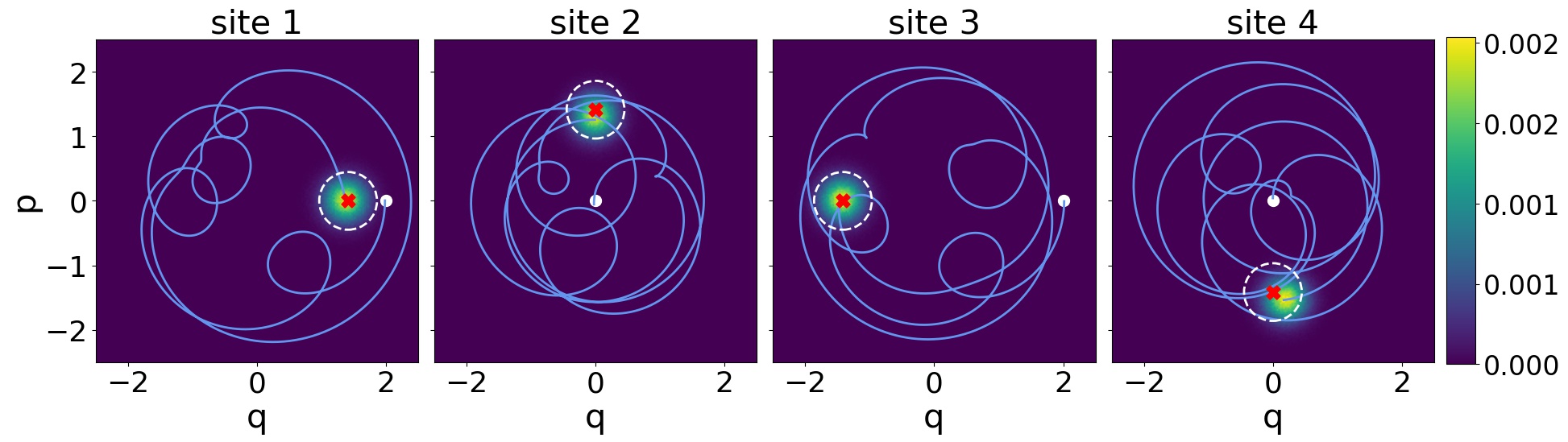}
	\end{subfigure}
	\caption{\textbf{Coherent targeting along chaotic control trajectory.} A coherent state (here for illustration $\heff = 1/10$ was chosen) is initialized centered on the density wave $\Vec{\phi}_{\alpha} = \sqrt{2/\heff}(1,0,1,0)$ (white dot). The inital phase space point is connected to the target $\Vec{\phi}_{\beta} = \sqrt{1/\heff}(1,i,-1,-i)$ (red cross) through a control trajectory found in the Bose-Hubbard mean-field limit (indicated as blue line). The $2\sigma$-curves of the initial and target state are represented with white dashed circles. A TWA simulation shows that driving the control system with the occupations of the control trajectory results in the coherent state evolving along the predesigned path, keeping its minimum uncertainty form. After the control time $t_c = 4.9\tau_{\text{DW}}$ the propagated state lands well inside the volume of the target state. }
	\label{fig:twaTargeting}
\end{figure}

Figure~\ref{fig:twaTargeting} shows the initial and final state of the protocol.  The full time evolution is provided in Animation 2 of the supplementary material.  In contrast to the spreading observed in Fig.~\ref{fig:BH_equilibration}, the coherent state keeps its minimum uncertainty form throughout the entire propagation, successfully arriving close the target mean-field. For completeness, a full quantum simulation of the time evolution is performed.  It gives the same results.  Figure~\ref{fig:orbitTargeting} shows that the expectation value of the occupations follow the classical mean-field solution perfectly in the control system, whereas they almost immediately deviate when evolving in the original interacting Bose-Hubbard system; see Fig.~\ref{fig:BH_equilibration}.


\subsection{Protocol robustness}
\label{subsec:errors}

Three mechanisms that might potentially degrade the quality of a particular application of coherent quantum targeting are: i) the proximity of initial and target state centroids from initial and final conditions of the control trajectory, respectively, ii) the precision with which the interactions can be made to vanish, and iii) the faithfulness with which the time-dependencies of the set of $\{\mu_j(t)\}$ match the occupancies of the control trajectory.  \change{These mechanisms are discussed below individually.  Note that an analysis of the latter item is presumably model dependent, and amongst other issues, might depend on the type of noise inherent in creating the $\{\mu_j(t)\}$.  Nevertheless, the effects of uncorrelated (white) noise in time are considered as a simple starting point.}

\subsubsection{Proximity}
\label{sec:prox}

\begin{figure}
	\centering
	\includegraphics[width = 0.8\textwidth]{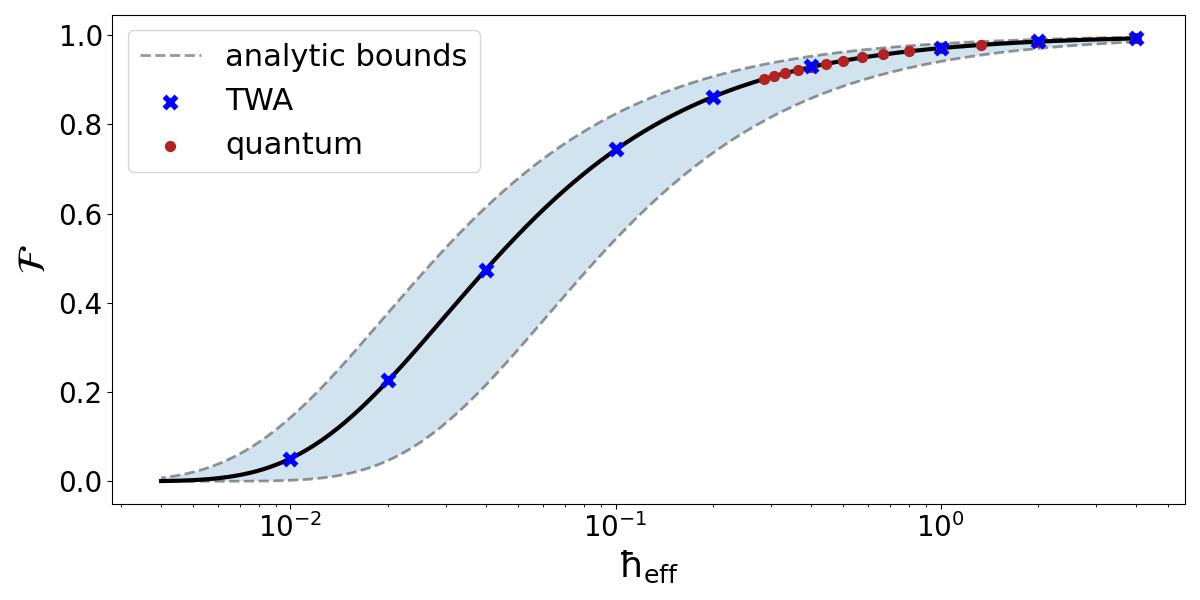}
	\caption{\textbf{Fidelity of the targeting protocol.} The protocol depends on the specific control trajectory used for targeting. The initial and final distance from the particular trajectory to the centroids of the respective states introduce a implicit $\heff$ dependence into the fidelity.  Upper and lower bounds depending on those distances are depicted. The exact fidelity (black curve) depends on the distance of the final states centroid to the actual target state and can be obtained numerically. The results obtained from both TWA and quantum simulations match the fidelity curve.  Note that the fidelity would remain unity for all $N$ if shift operators were additionally used to translate the initial and final coherent states.} 
	\label{fig:fidelityNoShifts}
\end{figure}

As mentioned at the end of Sec.~\ref{sec:background} and proved in \ref{sec:A2}, the specific quadratic nature of $\hH_c$ guarantees that a coherent state initially (exactly) centered on the control trajectory used to drive the system follows this trajectory perfectly without any spreading, and satisfies the associated time-dependent Schrödinger equation.  In addition, any coherent state centered anywhere remains a coherent state for all times, only they do not follow the control trajectory.  In general however, as mentioned at the beginning of this section, the initial and final conditions of any control trajectory, Sec.~\ref{subsec:targeting}, are slightly shifted from the initial and desired target states centroids, respectively.  Letting $t_c$ denote the time at which the control (heteroclinic) trajectory arrives closest to the target point, the shifts can be expressed as the norms
\begin{equation}
\delta x_{\text{initial}} = ||\Vec{x}_{\alpha} - \Vec{x}(0)||, \qquad \delta x_{\text{target}} = ||\Vec{x}_{\beta} - \Vec{x}(t_c)||\ ,
\end{equation}
resulting in a non-perfect overlap between the target state $\coh{\Vec{\phi}_{\beta}}$ and the actual time evolved final state 
\begin{equation}
\coh{\Vec{\phi}_\alpha(t_c)} \coloneqq \exp{-\frac{i}{\hbar}\int\limits_0^{t_c} dt \hH_{\text{c}}\left(t\right)}\coh{\Vec{\phi}_{\alpha}} \ .
\end{equation}
The shifts are minimized in the search for a control trajectory while simultaneously keeping the protocol time $t_c$ near its minimum.  Since this distance only has a meaning relative to the volume of the coherent state, an $\heff$ dependency gets introduced into the fidelity $\mathcal{F}$ of the control process.  States with larger filling factors and thus smaller volume are more affected by the shifts as shown in Fig.~\ref{fig:fidelityNoShifts}.  Based on the knowledge of $\delta x_{\text{initial}}$ and $\delta x_{\text{target}}$ bounds can be derived for the fidelity of the protocol:
\begin{equation}
\mathcal{F}_{\text{min}}  = \exp{-\frac{\left(\delta x_{\text{initial}} +  \delta x_{\text{target}}\right)^2}{2\heff}}, \quad \mathcal{F}_{\text{max}}  = \exp{-\frac{\left(\delta x_{\text{initial}} -  \delta x_{\text{target}}\right)^2}{2\heff}} \ .
\end{equation}
The exact overlap is given as
\begin{equation}
\mathcal{F} = \left|_{\text{coh}}\bra{\Vec{\phi}_{\beta}}e^{-\frac{i}{\hbar}\int\limits_0^{t_c} dt \hH_{\text{c}}\left(t\right)}\ket{\Vec{\phi}_{\alpha} }_{\text{coh}}\right|^2 = \exp{-\frac{\delta\Vec{x}^2}{2\heff}}\ ,
\end{equation}
where $\delta \Vec{x}$ is the distance between the centroids of the target and final state.  It could be derived using an adapted version of Heller's linearized wave packet dynamics~\cite{Heller75}, which would be rather involved due to the dependencies on the stability matrix.  Here, it is simpler to extract $\delta \Vec{x}$ numerically.  Interestingly, even though the control method is semiclassical in nature, the harmonic nature of the Hamiltonian causes the protocol to have almost perfect fidelity in the dilute quantum limit and then fall off for states with larger filling factor, i.e., inverse $\heff$.


\subsubsection{Residual interactions}
\label{sec:resint}

The second mechanism mentioned above that could potentially degrade the coherent quantum targeting is related to switching off the on-site interactions.  Despite the fact that there can be great experimental control over interactions~\cite{Chin10}, imagine nevertheless that the strength of the interaction cannot be \change{turned off} precisely, and there is a small perturbative residual interaction, which might also have some slow or weak time dependence.  Also, imagine that the residual strength can be measured or known by some means.  This would add a perturbative term to the control Hamiltonian, Eq.~\eqref{eq:conBH_H}, as follows:
\begin{align}
\hH_{\rm c}^{\rm \epsilon}\left(t\right) = - \sum_{j=1}^{L} \left(\had{j}\ha{j+1} + \had{j+1}\ha{j} - \mu_{j}(t)\hat{n}_j \right) + \frac{\epsilon(t)}{2} \sum_{j=1}^{L}\hat{n}_j\left(\hat{n}_j-1\right) \ .
\label{eq:CH_corrected2}    
\end{align}
The effect of such residual interactions would be twofold: the interactions would again lead to a deformation of the initially minimum uncertainty coherent states, and constructing the control Hamiltonian with $(\vec{q}(t),\vec{p}(t))_c$ would result in missing the desired target.
\begin{figure}
	\centering
	\includegraphics[width = \textwidth]{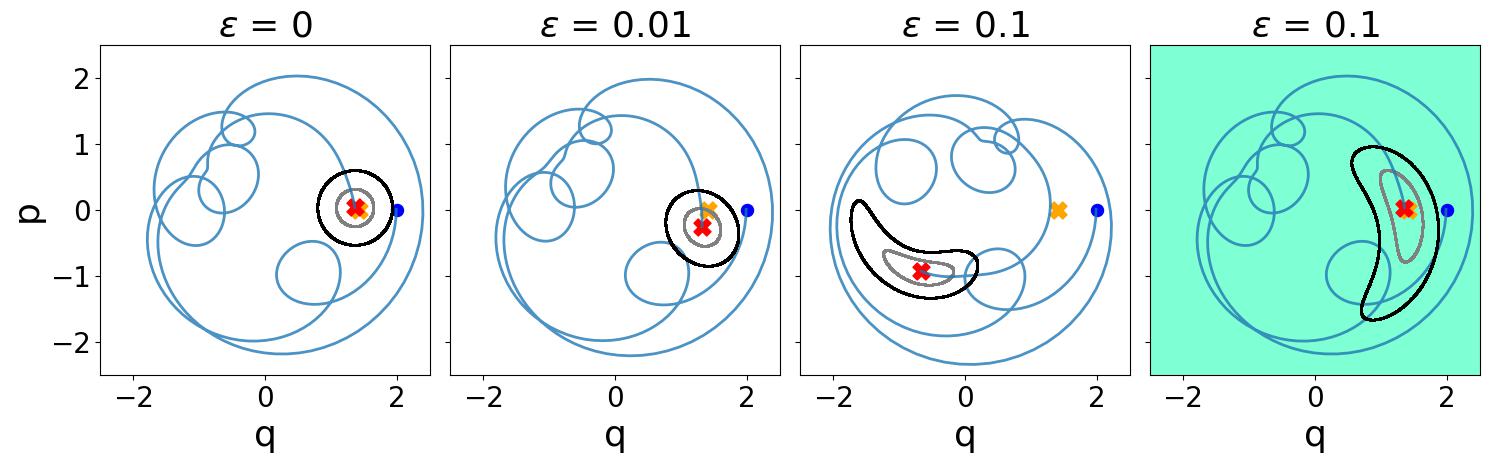}
	\caption{\textbf{Effect of constant weak residual interactions.} The trajectory no longer ends near the target and is no longer perfectly stable, which distorts the contours of the Wigner transform. Shown for only the first site, the final point of the control trajectory ends at the red cross, which is increasingly far from the target (orange cross) as $\epsilon$ increases.   The spreading also increases with $\epsilon$ as illustrated for the $1\sigma$-contour of the propagated Wigner density for $\heff=1$ (black) and $\heff=\change{1/4}$ (gray).  For very small $\epsilon$ $(= 0.01)$, both effects remain negligible. However for $\epsilon=0.1$, the control trajectory completely overshoots the target and there is a strong deformation of the coherent state.  Accounting for $\epsilon$ in the $\mu_j(t)$, the overshoot is corrected, however the spreading remains, as shown in the most right panel. The different state contour areas are a result of the depicted $1\sigma$-contour being only a cut through a higher-dimensional object. }
	\label{fig:effectsResidualInt}
\end{figure}
Both effects are illustrated in Fig.~\ref{fig:effectsResidualInt} for a constant residual interaction $\epsilon(t)=\epsilon$, as well as the improvement using the appropriately modified $\{\mu_j(t)\}$.  If the behavior of $\epsilon(t)$ is known, it is quite simple to take into  account.  Consider Hamiltonian equations, just as before in Eq.~\eqref{eq:conBH_eqM}, which were used to determine the $\{\mu_j(t)\}$, except with the inclusion of the $\epsilon$ term.  Now resolve the equations as before, the result is the replacement,
\begin{equation}
\mu_j(t) \Longrightarrow \mu_j(t) - \epsilon(t)
\label{eq:CH_corrected}
\end{equation}
in Eq.~\eqref{eq:CH_corrected2}.  It suffices to subtract $\epsilon(t)$ from the $\{\mu_j(t)\}$ to counteract the over or undershooting of the target. 

If $\epsilon$ is not compensated in the $\{\mu_j(t)\}$, the fidelity decay with $\epsilon$ is more significant as $\heff \rightarrow 0$. This indicates that the dominant error source is the perturbed control trajectory not arriving at the target.  It follows from the above argument that a non-zero distance of the final state to the target state affects more significantly the overlap of states with smaller volume (larger mean particle number).  If correcting the $\hH_c$ by adjusting the driving by the replacement, Eq.~(\ref{eq:CH_corrected}), an overall improved robustness to the residual interactions is obtained. Moreover, notably, the $\heff=1/4$ coherent state is less affected than the $\heff=1$ one.  The remaining error source is the deformation of the state, which is less significant for smaller volumes.
\begin{figure}
	\centering
	\includegraphics[width = 0.8\textwidth]{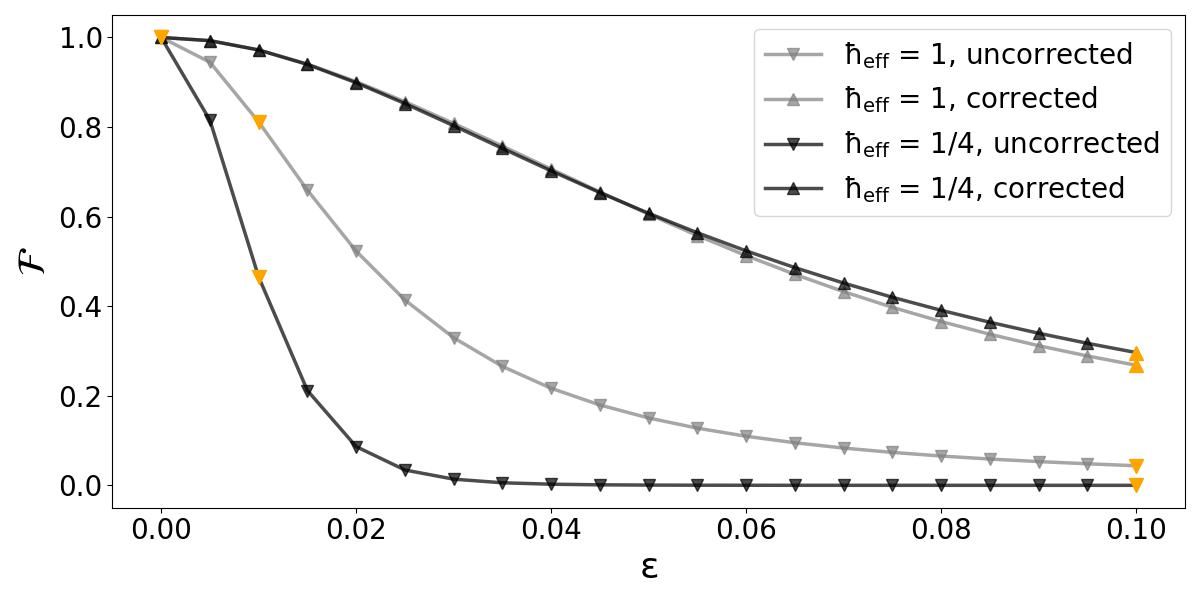}
	\caption{\textbf{Partial compensation of constant residual interactions.} For a constant $\epsilon(t)=\epsilon$, the fidelity of an uncompensated protocol quickly decays. The $\heff=1/4$ decays much faster than $\heff=1$.  The over or undershooting of the target is the dominant difficulty.  If the behavior of $\epsilon(t)$ is known, it can be accounted for following the replacement in Eq.~\eqref{eq:CH_corrected}. This leads to a significant overall improvement of the fidelity, even favoring the smaller $\heff$ case, due to the deformation being the remaining error. Since TWA is no longer exact with the interaction term present, the overlaps have been calculated in a full quantum simulation. The orange triangles mark the cases depicted in Fig.~\ref{fig:effectsResidualInt}.}
	\label{fig:residualInteractionQuantum}
\end{figure}

\change{
	\subsubsection{White noise} 
	\label{sec:whitenoise}
	
	The final potential mechanism that might degrade the fidelity of the control protocol has to do with how perfectly the $\mu_j(t)$ can be produced experimentally.  As a simple modeling starting point, imagine that there is a small amount of white (uncorrelated) noise generated in the $\mu_j(t)$.  Consider adding to each individual $\mu_j(t)$  white noise (uncorrelated over time and site index) of some given root mean square (RMS) magnitude.  This changes the mean field trajectory and its endpoint differs from the ideal.  As a function of the white noise strength (i.e., the RMS of a white noise realization), the RMS distance $\delta_{\rm RMS}$ between noisy and ideal trajectory endpoints are calculated for 50 realizations.  The results are shown in Fig.~\ref{fig::white noise}.
	\begin{figure}
		\centering
		\includegraphics[width=0.8\linewidth]{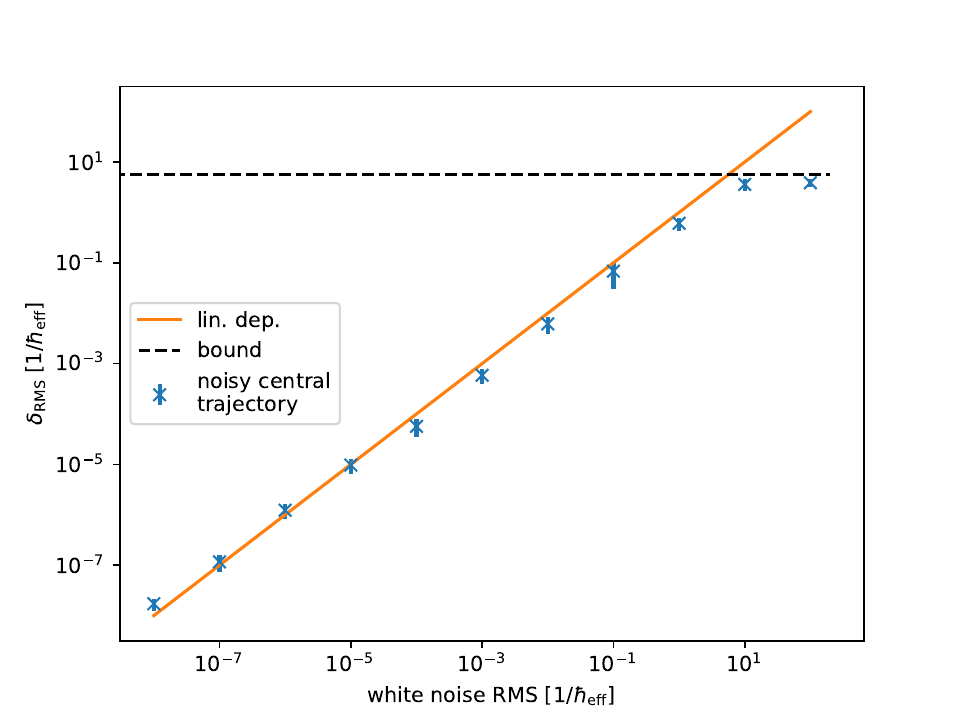}
		\caption{\change{\textbf{Final deviation of the target due to white noise.} An RMS average is calculated over 50 noise realizations for the final distance to the unperturbed trajectory. There is a linear dependency for low amplitudes due to the stability of the target protocol and cancellation of random changes. For strong noise magnitude (dominant with respect to the control $\mu_j(t)$), the final point is random and the distance is bounded by the phase space diameter. The errorbars display the run-to-run standard variance.}}
		\label{fig::white noise}
	\end{figure}
	
	It turns out that the RMS deviation depends linearly on the strength of the white noise up to a level beyond which it saturates.  The control trajectory is a stable solution and the perturbations are random with a vanishing mean.  This gives a perturbed trajectory that stays close to the ideal.  As the time evolution is integrated, there are significant cancellations of the uncorrelated deviations.  The trajectory with noise only slowly diffuses away from the ideal. 
	
	For perturbations comparable to the phase space size, i.e., occupations ($\sim 1/\hbar$), the actual control fields $\mu_j(t)$ are more or less negligible and the motion is random, but nevertheless the deviation to the target point is bounded by the phase space size. Therefore, the curve saturates for strong enough white noise.}

\subsection{Stabilizing periodic trajectories}\label{subsec:periodic}

Nearly forty years ago, the concept of ``quantum scarring of eigenstates'' was introduced by Heller~\cite{Heller84}.  He argued for and observed excess intensity in the neighborhood of short unstable periodic trajectories in the quantum eigenstates of a strongly chaotic dynamical system.  The critical point was that there were expectations that the eigenstates would behave similarly to random wave functions subject only to energy surface constraints~\cite{Berry77, Voros79, McDonald79, McDonaldThesis}, i.e., $\delta(E-H)$ must somehow be respected.  Furthermore, there were quantum ergodic theorems regarding equidistribution and expectation values of smooth operators giving classical values for nearly all eigenstates~\cite{Shnirelman74}.  The excess intensity observed can be considered a weak form of dynamical localization, and yet somehow must be consistent with the mathematical ergodic theorems, which is a key element of the surprise of scarring's existence.

In the many-body context, the Eigenstate Thermalization Hypothesis is the closest analog of the quantum ergodicity discussed above~\cite{Deutsch91, Srednicki94}. Unexpected deviations from this hypothesis in certain cases might therefore provide the notion of a many-body quantum scarring analogous to Heller's quantum scarring. Experimental effects seen in a 51-atom Rydberg chain, and recently even in tilted Bose-Hubbard systems, were introduced as many-body scarring~\cite{Su23,Bernien17, Turner18, Serbyn21}, although it is not necessarily clear that it is conceptually the same as in the original context. In fact, an emergent local regularity of the dynamics, at least on short to medium time scales, may underlie this many-body scarring phenomenon~\cite{Khemani19}.  Thus, in some cases many-body quantum scarring may be a result of at least some partial local integrable or near-integrable dynamics.  Nevertheless, there exists at least one example in the many-body context in which the explicit connection to unstable periodic mean-field solutions has been made~\cite{Hummel23}.  With the keen current interest in many-body periodic or partially periodic dynamics~\change{\cite{Ljubotina22, Ljubotina24}}, it is worth looking at how optimal quantum coherent control can be utilized to stabilize periodic dynamics in a many-body system.  The method is applied to simpler symmetric and much more complicated asymmetric unstable trajectories next.

\subsubsection{Symmetric perodic mean-field control trajectory}

\begin{figure}
	\centering
	\includegraphics[width = 0.8\textwidth]{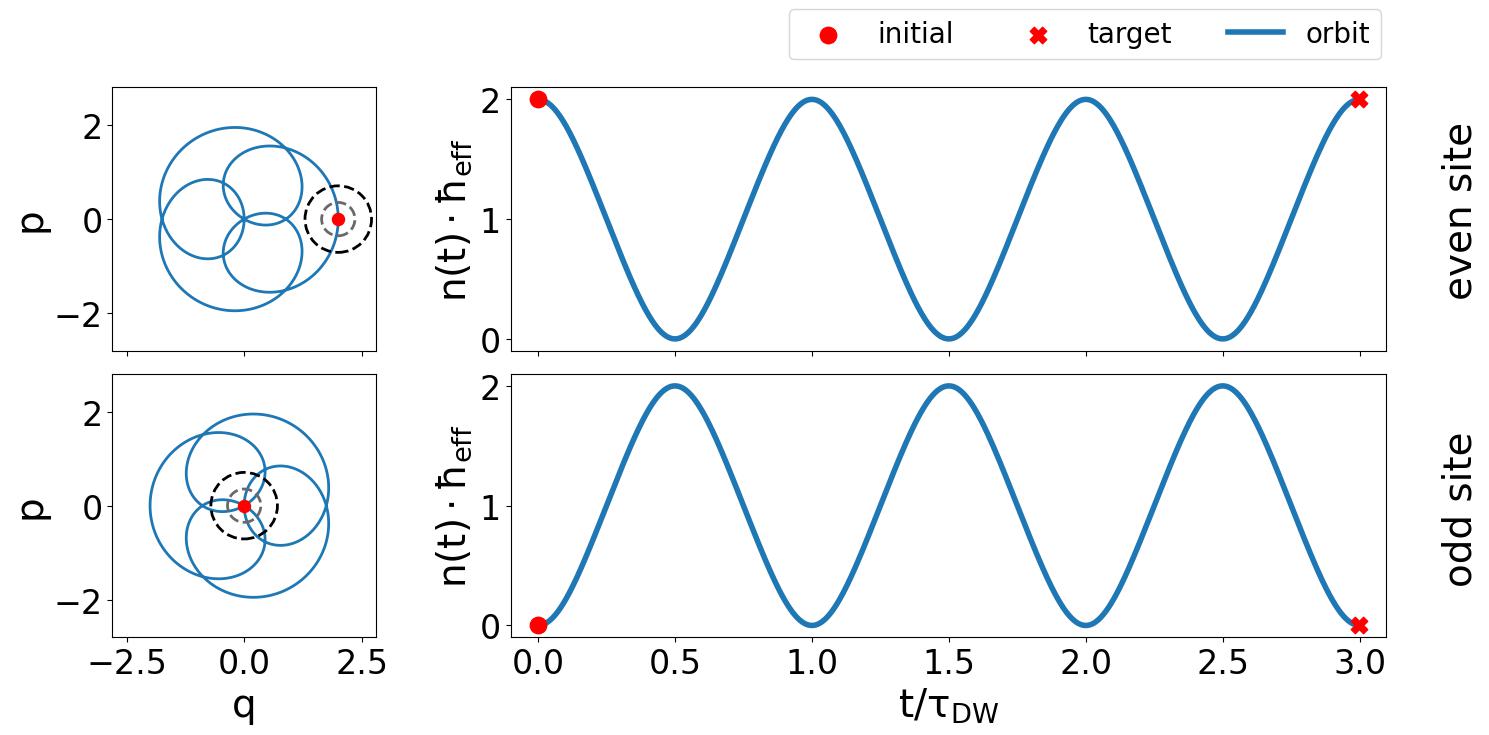}
	\caption{\textbf{Symmetric periodic trajectory.} For $\gamma = 2.0607$ there is a periodic mean-field solution very close to the density wave that is symmetric for all even and odd sites. After three periods of the occupations (or three driving periods) the solution returns to its original phase. }
	\label{fig:pt}
\end{figure}

In Sec.~\ref{sec:dw}, it is mentioned that the density wave exhibits periodic population inversion.  This does not constitute a periodic trajectory unless the phase is also periodic at the same time.  It turns out that there are  truly periodic trajectories very close to the density wave initial condition for certain values of $\gamma$.  One of these values is selected and illustrated in Fig.~\ref{fig:pt}.  For each cycle of population inversion, 
the phase advances by $2\pi/3$, which after the third cycle completes the periodic trajectory.  Note the simple appearance, similar to a Lissajous figure, of the quadrature variables for the trajectory.  In essence, the initially empty site is $\pi/2$ out of phase with the occupied site.  There is an Animation 3 in the supplementary material illustrating the continuous time evolution. 

\begin{figure}
	\centering
	\includegraphics[width = \textwidth]{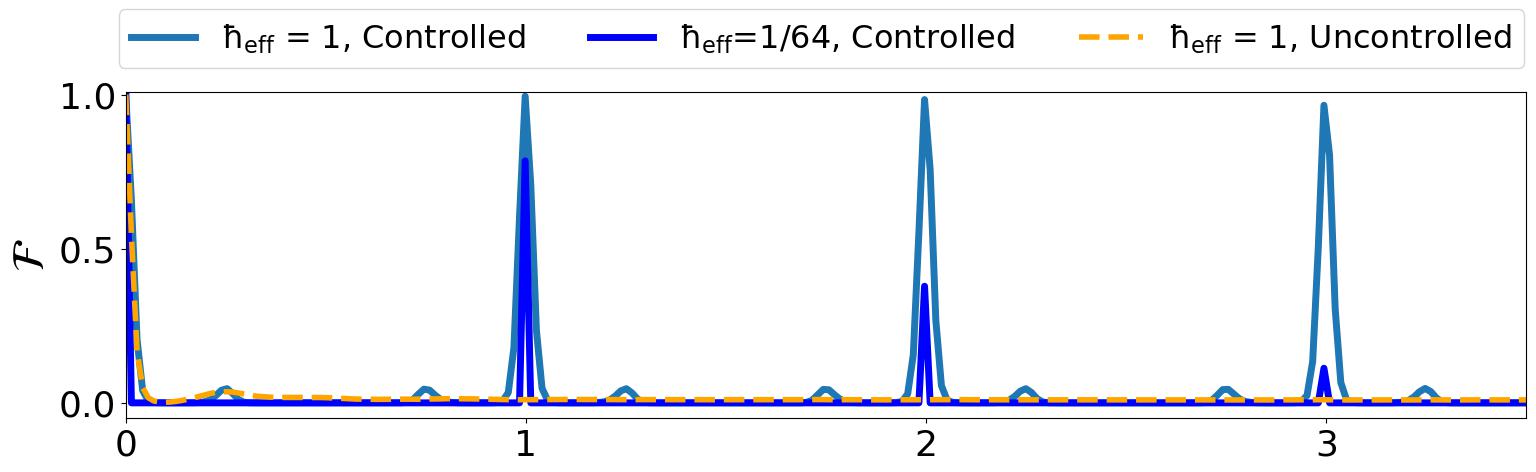}
	\caption{\textbf{Quantum recurrences in the controlled system.} When propagated in the uncontrolled system, no recurrences appear. For $\heff = 1$, the recurrences are almost perfect due to the large size of the state in phase space. In addition, smaller side peaks are visible because the volume is so large. Those disappear for the smaller inherent volume for the $\heff = 1/64$ case, together with more imperfect recurrences. Naturally, the smaller volume of the $\heff = 1/64$ case also generates a sharper peak.}
	\label{fig:perfid}
\end{figure}

For the same reasons that the fidelity of the control protocol is not perfect in Fig.~\ref{fig:fidelityNoShifts}, i.e., the trajectory initial conditions do not perfectly match the initial state centroids, the achievement of perfect periodicity degrades with increasing time and decreasing $\heff$.  This is illustrated in Fig.~\ref{fig:perfid}
for the two cases $\heff=1, 1/64$ up to three periods of the periodic motion.  The tall peaks indicate the initial state is recurring at integer multiples of the periodic control trajectory with the $\heff=1$ case falling off slowly with increased time, and the $\heff=1/64$ case falling off much quicker.  Perfect recurrences would be restored by an initial shift of the initial state centroids to the initial conditions of the periodic mean-field trajectory.


\subsubsection{Asymmetric periodic mean-field control trajectory}

\begin{figure}
	\centering
	\includegraphics[width = 0.8\textwidth]{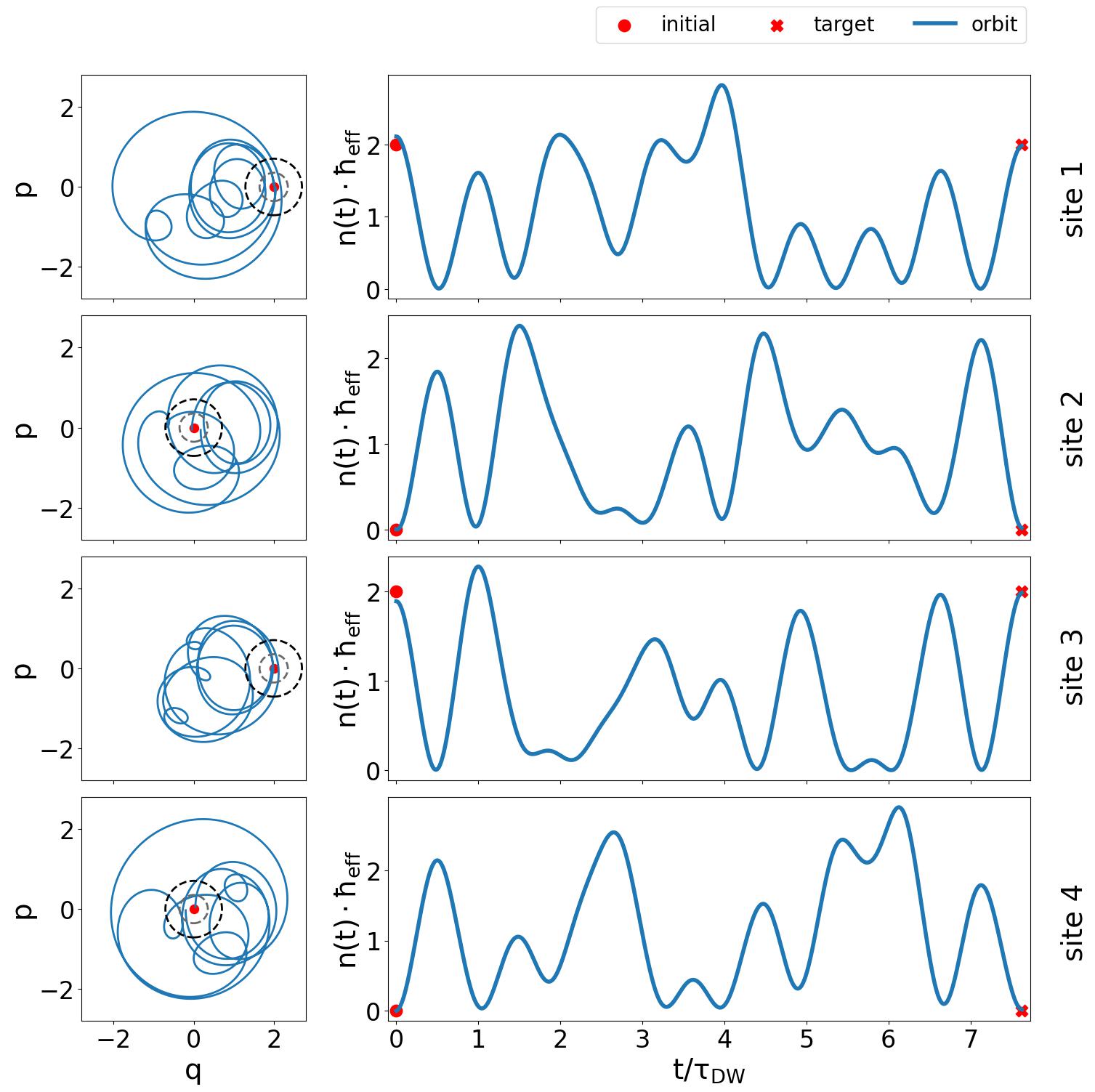}
	\caption{\textbf{Asymmetric periodic trajectory.} There exist also a dense set of highly nontrivial periodic mean-field trajectories, which could serve as control trajectories. Here, a trajectory is shown for $\gamma = 1.3$.  Its evolution is extremely close to a periodic mean-field trajectory. Due to technical complications, the exact periodic trajectory's initial conditions are not sought to greater precision, and in any case, the density wave is not perfectly centered on the control trajectory either.  }
	\label{fig:asympertra}
\end{figure}

Although more challenging, it is possible to locate periodic mean-field trajectories with far less symmetry that are in the close neighborhood of a density wave.  An example is shown in Fig.~\ref{fig:asympertra}.  In either the quadrature variables, which do not even faintly resemble a Lissajous figure, or even just the occupancies, the control trajectory follows a highly nontrivial evolution before it returns.  The complete time evolution can again be found for this case as Animation 4 in the supplementary material.  The quality of the recurrence for this longer, more complicated trajectory, is not quite as good as for the symmetric trajectory in the previous subsection.

\section{Coherent quantum targeting in $2D$ lattices}
\label{sec:2d_applications}

A considerable utility of cold atom optical lattices lies in their capacity to be created in any number of dimensions. A priori, the control protocol is not limited to $1D$ systems, however higher-dimensional systems lead to exponentially expanding search spaces for heteroclinic trajectories.  Thus, to target arbitrary states requires sophisticated search algorithms~\cite{Altman18, Tomsovic18b}, and even so it may become effectively impossible.  Nevertheless, there do exist certain classes of symmetric, periodic lattice configurations for which it is possible to map the mean-field dynamics to that of $1D$ periodic rings~\cite{Steinhuber20}. In this way, the protocol can be reduced to identifying the corresponding trajectory in the lower-dimensional space, which renders the search feasible.  Here, $2D$ lattice examples are treated in which the mean-field dynamics are explicitly mapped to that of a $1D$ ring, reducing the problem again to finding control trajectories in a numerically accessible search space.

There is more than one way to generate an appropriate mapping for these purposes.  For the one relied on below, the sites are mapped into higher-dimensional lattices such that the nearest-neighbor associations are preserved.  There is a change in nearest-neighbor multiplicity with increasing dimension that requires a renormalization of the hopping parameter, which once taken into account generates a dynamics in the larger lattice that appears as multiple copies of the $1D$ ring used in the mapping~\cite{Steinhuber20}.  As long as the initial and target states respect the implied discrete symmetries, the optimal coherent quantum targeting in higher-dimensional lattices reduces to again identifying optimal control trajectories in the $1D$ ring.

\begin{figure}[t]
	\centering
	\begin{subfigure}{0.32\textwidth}
		\centering
		\includegraphics[width = 0.95\textwidth]{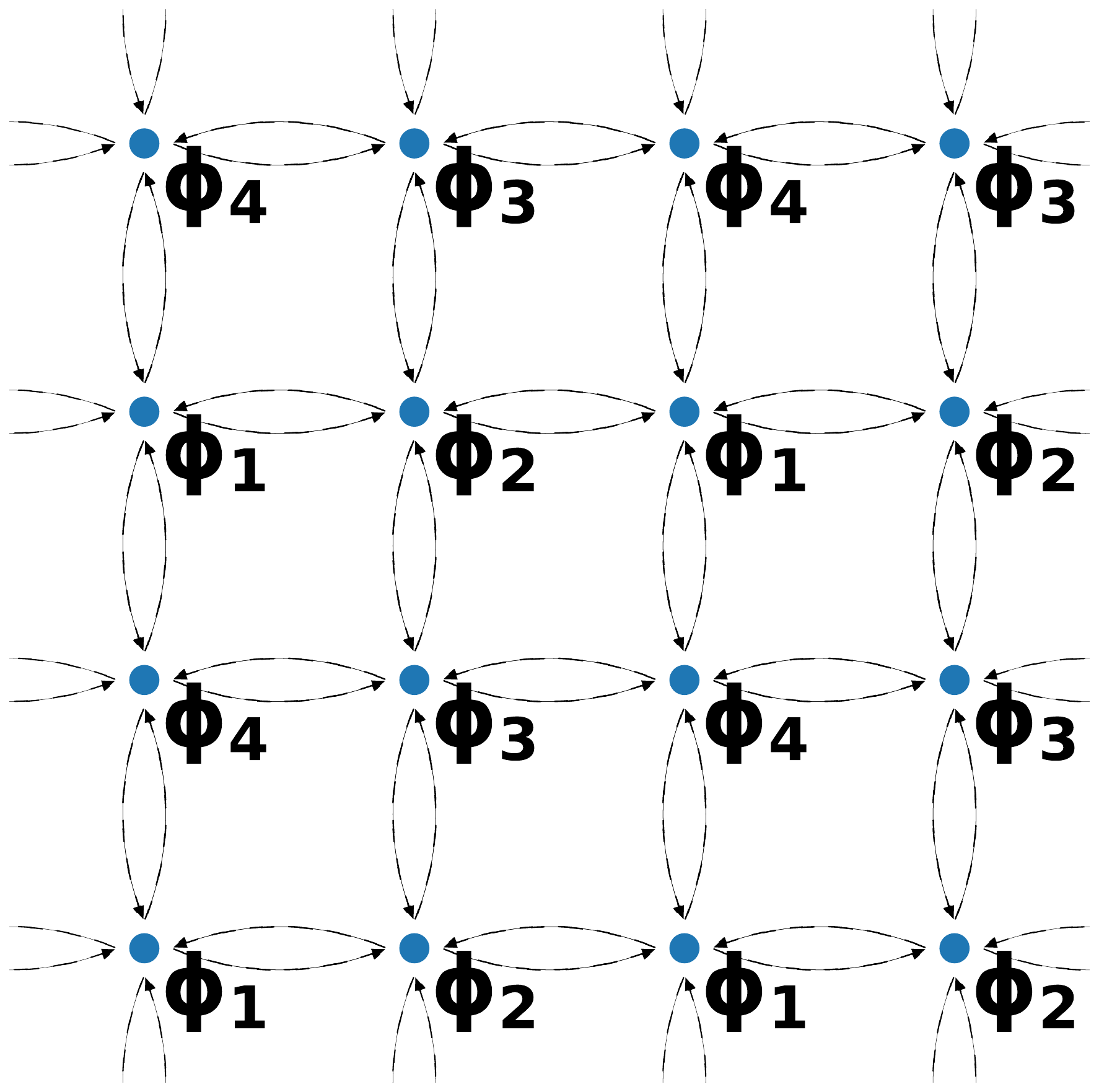}
		\caption{}
		\label{fig::2dVortex}
	\end{subfigure}
	\begin{subfigure}{0.32\textwidth}
		\centering
		\includegraphics[width = 0.95\textwidth]{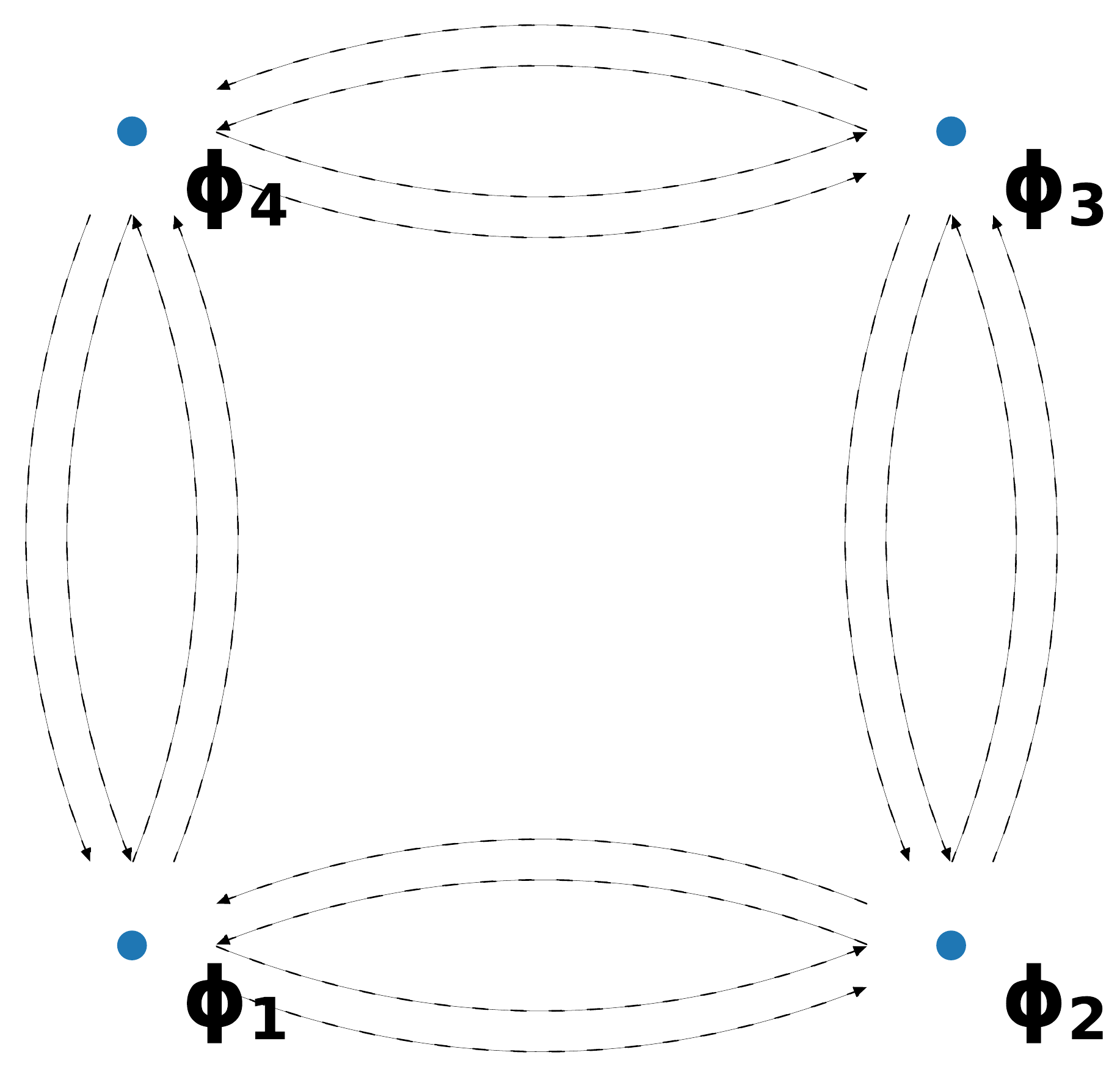}
		\caption{}
	\end{subfigure}
	\begin{subfigure}{0.32\textwidth}
		\centering
		\includegraphics[width = 0.95\textwidth]{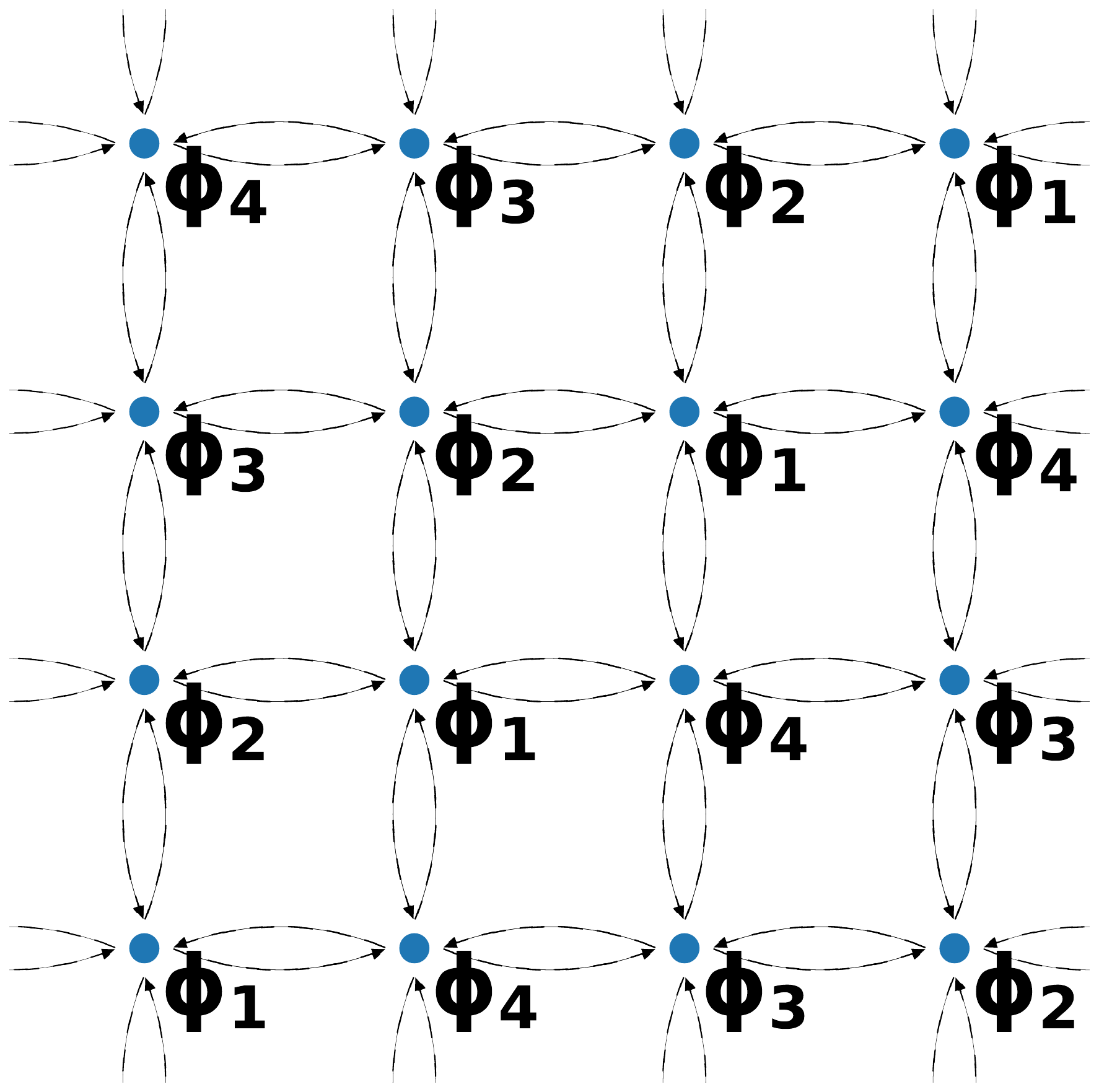}
		\caption{}
		\label{fig::2dshifted}
	\end{subfigure}
	\caption{\textbf{Two possible schematic mappings of $2D$ mean-field dynamics to a $1D$ ring.}  By initializing the condensates on a $2D$ periodic lattice in the arrangement shown in panels (a) or (c), the mean-field dynamics of each lattice site can be mapped to the dynamics of the corresponding site on a $1D$ ring, (b), with double the hopping parameter, i.e.~$J_{\text{1D}} = 2J_{\text{2D}}$. \change{In greater detail, a 1D mean-field solution $\Vec{\phi}^{1D} = (\phi_{1},\phi_{2},\phi_{3},\phi_{4}))$ with a 1D Hamiltonian $H_{1D}(2J,U,\Vec{\mu}/2)$ is mapped via the patterns (a) or (c) to a 2D mean-field solution with a 2D Hamiltonian $H_{2D}(J,U,\Vec{\mu})$}.  
	}
	\label{fig::schematicMapping}
\end{figure}

Following this particular prescription, two possible lattice arrangements are shown in Fig.~\ref{fig::schematicMapping}, that allow for the dynamics 
of a periodic $4 \times 4$ lattice to be mapped onto the dynamics of a four-site ring.  In fact, the mappings work for any $4n \times 4m$ $2D$  lattice with periodic boundary conditions (including $n \rightarrow \infty$). Here the mean-field dynamics reduce to the four-site ring with double the hopping parameter $J_{\text{1D}} = 2J_{\text{2D}}$~\cite{Steinhuber20}.  The two examples given are a lattice consisting of four-site ring building blocks or of shifted stripes, respectively. By initializing the lattice in one of those symmetric ways, it is possible to guide a many-body state along a control trajectory by driving the $2D$-lattice system with its mean-field occupations, $\mu_{\Vec{R}}(t)$, corresponding to those of the reduced $1D$ system, $\Vec{\mu}(t)$. The $2D$-lattice control Hamiltonian,
\begin{align}
\hH_{\text{c}}^{\text{2D}}\left(t\right) = -\sum_{\left<\Vec{R}, \Vec{R}' \right>} \big( \had{\Vec{R}}\ha{\Vec{R}'} + \text{h.c.} \big) + \sum_{\Vec{R}}
\mu_{\Vec{R}}(t)\hat{n}_{\Vec{R}},
\label{eq:2dconBH}
\end{align}
is analogous to the $1D$ case after accounting for the doubled number of neighboring sites in the chemical potentials. 

\begin{figure}[t]
	\centering
	\begin{subfigure}{0.49\textwidth}
		\centering
		\includegraphics[width = 0.8\textwidth]{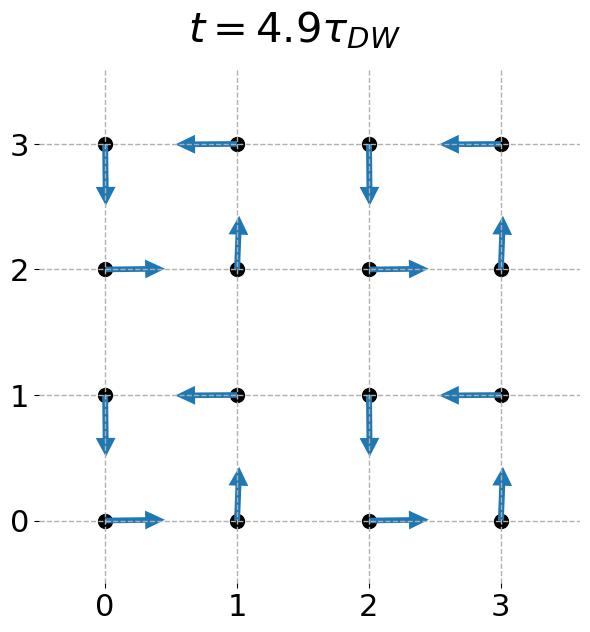}
		\caption{}
	\end{subfigure}
	\begin{subfigure}{0.49\textwidth}
		\centering
		\includegraphics[width = 0.8\textwidth]{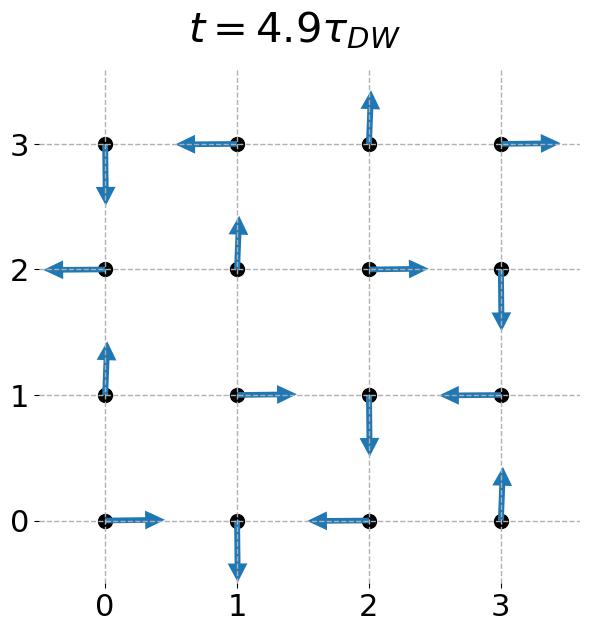}
		\caption{}
	\end{subfigure}
	\caption{\textbf{Targeting in $2D$ lattices.} Mapping the mean-field dynamics of a symmetrically arranged $2D$ lattice to a four-site ring allows for targeting certain classes of states by identifying their respective lower-dimensional control trajectory. By initializing the lattice using the arrangements presented in Fig.~\ref{fig::2dVortex} or Fig.~\ref{fig::2dshifted} with the initial conditions of the control trajectories previously discussed and propagating the mean-field dynamics, it is possible to prepare the homogeneous lattice states shown in (a) and (b), respectively. The occupation of each site is encoded in the arrow length and the phases in the angle of the corresponding arrows read clockwise from the right horizontal.}
	\label{fig::2dTargeting}
\end{figure}

The two target configurations shown in Fig.~\ref{fig::2dTargeting} result from the mappings of Fig.~\ref{fig::schematicMapping} applied to the $1D$-ring control trajectory of Sec.~\ref{subsec:targeting}.  The initial condition consists of alternating occupied and unoccupied sites satisfying either mapping. The full evolutions are shown in the supplementary material, Animation 5 and Animation 6, respectively, where the sites evolve according to the symmetry of the mappings.  This way it is possible, for example, to target the homogeneous lattice state where the respective condensate phases form a vortex structure as shown in Fig.~\ref{fig::2dTargeting}a or a shifted stripe structure as shown in Fig.~\ref{fig::2dTargeting}b.

Applying the same logic it is possible to also create periodic many-body states on a $2D$ lattice by for example using the periodic trajectories discussed in Sec.~\ref{subsec:periodic}.  The entire time evolution is again presented in the supplementary material, Animation 7 and Animation 8, respectively.  Naturally there also exist multiple mappings that reduce the dynamics of a $3D$ lattice to that of a four-site ring with adjusted hopping parameter $J_{\text{1D}} = 3J_{\text{3D}}$.


\section{Conclusion}
\label{sec:conc}

The presence of chaos poses fundamental challenges for controlling the evolution of a classical dynamical system, nevertheless it is long known that it is possible to convert the inherent difficulties caused by instability and ergodicity into a resource~\cite{Ott90}.  Recently, it was shown that with respect to quantum systems, an additional requirement is needed, i.e.~the suppression of quantum state spreading~\cite{Tomsovic23}.  Done originally in a Schr\"odinger or single particle context, a feasible extension into the many-body domain, here for ultracold atoms in an optical lattice, requires the identification of a time-dependent control Hamiltonian, and in this way build a quantum simulator controlling many-body quantum chaos.

The extension of chaotic classical targeting to coherent quantum targeting is facilitated through the use of semiclassical methods in which trajectories in the mean-field/classical limit and their stability analysis provide the basis for determining the necessary perturbations or time-dependent modifications of the system.  The quadrature phase space formulation of a many-body system of ultracold bosons results in a mean field limit, i.e.~large filling factor, of ordinary Hamiltonian dynamics.  Thus, the adaptation to a many-body system parallels closely the prior work in a Schr\"odinger dynamics context.  In particular, the same class of trajectories, heteroclinic ones, provides the time-dependent chemical potentials that determine the control Hamiltonian.

As in~\cite{Tomsovic23,Tomsovic23b}, the protocol is designed for guiding localized quantum states.  All of the results are shown for minimum uncertainty coherent states, but they apply immediately to the wider and important class that includes number projected coherent states. To control the spreading of the many-body state a control system is introduced that is essentially a time-dependent harmonic oscillator driven with the on-site occupations along the specific control trajectory. In practice this means that interactions between the bosons is turned off in an experiment, which is something that in optical lattice systems can usually be achieved to high precision, \change{e.g. using Feshbach resonances as a basis for one possible method.} Together with the fact that the protocol reduces to controlling the chemical potentials of the individual sites in a smooth, time-dependent fashion we believe this protocol to be suitable for practical experimental implementation.

Due to the harmonic nature of the system and the $U(1)$ symmetry of the dynamics any (number projected) coherent state placed on the control trajectory used to drive the Hamiltonian satisfies the time evolution of the system exactly. This means that the protocol provides a tool to propagate a broad range of states from arbitrary initial and target points in quadratures given that the underlying dynamics is sufficiently ergodic, opening up new possibilities in state preparation for optical lattices.  An example is shown of a relatively short trajectory that can be used to target a homegeneous lattice state with well defined, non-trivial phases between the condensates.

There are a few aspects of quantum targeting that may impact any assessment of how robust it can be against imperfections.  First, the ideal would be to make the tiny shifts of the initial state centroids to the initial conditions of the optimal control (heteroclinic) trajectories.  Likewise, the final state centroids on those trajectory endpoints would be slightly shifted to the target state's values.  In fact, the initial shift is the single most important element in classical targeting, whereas suppressing the state spreading is the critical element in the quantum case.  By virtue of the uncertainty principle, the optimal trajectory is already well represented inside the quantum state's Wigner transform.  By not performing the shifts, a decrease of the fidelity is introduced into the protocol, which gets worse as the filling factor increases.  This is observed numerically by calculating the overlap between the propagated and desired target state. 

A second imperfection would be due to the inability to turn off the interactions exactly, which left uncompensated quickly breaks down the control protocol.  Although, it is anticipated that the on-site interaction strength can, in many circumstances, be controlled experimentally to high precision, if any non-vanishing residual interaction can be characterized, then a simple rectification expressed in Eq.~\eqref{eq:CH_corrected} solves the over or undershooting effect, which would be the main contributor to a fidelity decay.  The local deformation of the initially circular Wigner density contours at the target state remains, but improves with increasing filling factor.  A third issue, that of imperfect control over the $\{\mu_j(t)\}$, \change{may} require appropriate modeling of noise or other imperfections \change{in a particular physical setup}, which is left for future consideration. \change{However, in the case of white noise (uncorrelated fluctuations in the individual chemical potentials over time), the strength of this type of noise has a linear dependence on the distance deviation in the targeting protocol and the control protocol is not particularly sensitive.}

For the examples given, it is necessary to introduce chemical potential offsets in order to place an initial state and a target state on the same energy surface such that chaotic trajectories can provide transport paths between them. If it were desired not to have these offsets at the end, although not discussed here, it would be possible to consider say adiabatically turning off the offsets and seeking some modified chaotic transport path between the initial and target states.  For example, the momentum target state of Sec.~\ref{sec:1d_applications} would become a fixed state of the system without interactions and chemical potential offsets, which might be a desired goal as well.  Additional time-dependence in the chemical potential offsets does not add much, if any, additional complications with respect to solving Hamilton's equations, but it adds an extra dimension to the complexity of the search for an optimal control trajectory.  This issue of possible unwanted final chemical potential offsets does not arise if the initial and targets states are identical, as for the periodic trajectory cases, as they are automatically on the same energy surface without offsets.

The presented $1D$ examples are of greater utility and versatility than they might appear at first sight.  By mapping the sites into higher-dimensional lattices such that the nearest-neighbor associations are preserved, it is possible to control much larger and higher-dimensional lattices without suffering from the exponential growth of search volumes in phase space.  For initial and target states that respect certain symmetric lattice configurations, optimal coherent quantum targeting in higher-dimensional lattices is no more complicated than finding control trajectories in the $1D$ ring to which the mean-field dynamics can be mapped. On the other hand, for arbitrary lattice configurations in high dimensions, a systematic search for control trajectories becomes  exponentially challenging with system size.

There are a number of interesting directions that future optimal coherent quantum targeting research might follow from the technical/pratical side to more general extensions.  This includes extending the methods to spin chains and fermionic systems, which present several new challenges.  It might be advantageous to replace the current methods of identifying control trajectories with machine learning methods.  Additionally, more work is needed to understand how the exponential speed up of heteroclinic pathways relates to quantum speed limits.  Finally, a direction worth mentioning is the investigation of many-body interference using the control protocol.


\subsection*{Acknowledgments}
We thank Monika Aidelsburger for useful conversations.  We acknowledge support by the Deutsche Forschungsgemeinschaft (DFG, German Research Foundation), project Ri681/15-1, within the Reinhart-Koselleck Programme and Vielberth Foundation for financial support. MS is funded through a fellowship by the {\em Studienstiftung des Deutschen Volkes}.
\clearpage

\appendix
\section{A formulation of quantum control from the perspective of auxiliary fields}
\label{sec:A1}

The purpose of this appendix is to provide a picture of the control protocol discussed in the main text from the point of view of the semiclassical approximation implemented within the auxiliary field formulation of interacting field theories.  The auxiliary field method, see for example~\cite{Negele98}, is an exact representation of the path integral form of the propagator where interactions are accounted for by the response of the system to all possible time-dependent external potentials. It is based on the functional identity 
\begin{equation}
\label{eq:HS}
\int \frac{{\cal D}[x]}{Z(M)}{\rm e}^{\frac{i}{2}\int_{0}^{t}ds \int_{0}^{t}ds' x(s)M(s,s') x(s')}{\rm e}^{i\int_{0}^{t}dsx(s)y(s)}={\rm e}^{\frac{i}{2}\int_{0}^{t}ds \int_{0}^{t}ds' y(s)(M^{-1})(s,s') y(s')}
\end{equation}
also known in the field theory literature as the Hubbard-Stratanovich (HS) transformation. Here $M(s,s')$ is a kernel with a functional inverse defined by
\begin{equation}
\int_{0}^{t} ds' M(s,s')(M^{-1})(s',s'')=\delta(s-s''),
\end{equation}
and $Z(M)$ is a normalization constant defined by comparing both sides of Eq.~(\ref{eq:HS}) with $y(s)=0$.  The generalization for multi-component fields, ($\vec{x}(s),\vec{y}(s))$, is straightforward by also introducing a matrix structure $M(s,s') \to {\bf M}(s,s')$.

To apply the HS transformation to the case of interest here, start with the coherent state form of the propagator for interacting bosonic fields, 
\begin{equation}
K(\vphi_{\beta},\vphi_{\alpha};t):=\langle \vphi_{\beta}|{\rm e}^{-\frac{i}{\hbar}\hat{H}t}|\vphi_{\alpha}\rangle 
\end{equation}
with $\hat{H}$ given by Eq.~(\ref{eq:BH_H}).  This propagator admits a path integral representation, found using standard techniques generically given by
\begin{equation}
\label{eq:prop}
K(\vphi_{\beta},\vphi_{\alpha};t)=\int{\cal D}[\vphi,\vphi^{*}] {\rm e}^{i (R_{\rm free}[\vphi,\vphi^{*}]+R_{\rm int}[\vphi,\vphi^{*}])},
\end{equation}
and defined by the non-interacting (free) $R_{\rm free}$ and interacting $R_{\rm int}$ terms in the action. As a rule, the free term contains functionals that are at most quadratic in the complex fields $\phi_{j}(s)$. Interaction terms, for the case of particular interest here, are in turn given as linear combinations of terms bilinear in the local occupations
\begin{equation}
\label{eq:Int}
\hat{V}_{\rm int}=\frac{1}{2}\sum_{i,j}v_{ij}\hat{n}_{i}\left(\hat{n}_{j}-\delta_{ij}\right) {\rm \ \ \ \ with \ \ \ } \hat{n}_{i}=\hat{a}^{\dagger}_{i}\hat{a}_{i}, {\rm \ \ and \ }v_{ij}=v_{ji}.
\end{equation}
Therefore, neglecting issues of ordering in defining the classical action functional (see~\cite{Wilson11} and~\cite{Bruckmann18} for details)
\begin{equation}
\label{eq:Rint}
R_{\rm int}[\vphi,\vphi^{*}]=\frac{1}{2}\int_{0}^{t}ds \sum_{i,j}v_{ij} |\phi_{i}(s)|^{2} \left(|\phi_{j}(s)|^{2}-\delta_{ij}\right).
\end{equation}

As is well known, the reason why the path integral, Eq.~(\ref{eq:prop}), cannot be evaluated in closed form is that the interaction term is not of Gaussian type. However, by means of the HS transformation, the fourth-order terms, Eq.~(\ref{eq:Rint}), can be decoupled in favor of the (vector) real auxiliary field $\vec{\sigma}(s)$ obtaining
\begin{equation}
K(\vphi_{\beta},\vphi_{\alpha};t)=\int \frac{{\cal D}[\vec{\sigma}]}{Z({\bf v})}{\rm e}^{\frac{i}{2}\int_{0}^{t}ds\vec{\sigma}(s){\bf v}^{-1}\vec{\sigma}(s)}K_{\vec{\sigma}}(\vphi_{\beta},\vphi_{\alpha};t)
\end{equation}
where the reduced propagator  
\begin{equation}
\label{eq:Ksigma}
K_{\vec{\sigma}}(\vphi_{\beta},\vphi_{\alpha};t)=\int{\cal D}[\vphi,\vphi^{*}] {\rm e}^{i (R_{\rm free}[\vphi,\vphi^{*}]+\int_{0}^{t}ds\sum_{j}\sigma_{j}(s) n_{j}(s))}, {\rm \  \ with \ }n_{j}(s)= |\phi_{j}(s)|^{2},
\end{equation}
now has the key property of representing a free field evolving under the time-dependent external field $\sigma_{j}(s)$ coupled at the $j$-th site to the local occupations, i.e., a time-dependent local chemical potential.

Next, Gaussian path integrals, even with time-dependent parameters as in Eq.~(\ref{eq:Ksigma}) preventing us to find their solution in closed form, can be exactly expressed by means of the solution of the corresponding classical (mean-field) problem, which in this case is unique.  In other words,
\begin{equation}
\label{eq:Ksex}
K_{\vec{\sigma}}(\vphi_{\beta},\vphi_{\alpha};t)=A(t,[\vec{\sigma}]){\rm e}^{i (R_{\rm free}^{\rm mf}(\vphi_{\beta},\vphi_{\alpha},[\vec{\sigma}])+\int_{0}^{t}ds\sum_{j}\sigma_{j}(s) n_{j}^{\rm mf}(s,\vphi_{\beta},\vphi_{\alpha},[\vec{\sigma}])))}
\end{equation}
where $R^{\rm mf},n_{j}^{\rm mf}$ are now functionals of the auxiliary fields $\vec{\sigma}$, as indicated by $[.]$, and functions of $t,\vphi_{\beta},\vphi_{\alpha}$. These objects are obtained by
\begin{eqnarray}
\label{eq:Rmf}
&&R_{\rm free}^{\rm mf}(\vphi_{\beta},\vphi_{\alpha},t,[\vec{\sigma}]) \\ &=&R_{\rm free}[\vphi(s)=\vphi^{\rm mf}(s,\vphi_{\beta},\vphi_{\alpha},t,[\vec{\sigma}]),\vphi^{*}(s)=\vphi^{\rm mf}(s,\vphi_{\beta},\vphi_{\alpha},t,[\vec{\sigma}])^{*}] \nonumber
\end{eqnarray}
and 
\begin{equation}
\label{eq:nmf}
n_{j}^{\rm mf}(s,\vphi_{\beta},\vphi_{\alpha},t,[\vec{\sigma}])=|\phi_{j}^{\rm mf}(s,\vphi_{\beta},\vphi_{\alpha},t,[\vec{\sigma}])|^{2}  
\end{equation}
from the unique solution $\vphi^{\rm mf}(s,\vphi_{\beta},\vphi_{\alpha},t,[\vec{\sigma}])$ of the Euler-Lagrange (mean field) equations
\begin{equation}
\label{eq:EL}
\frac{\delta}{\delta \vphi^{*}}\left(R_{\rm free}[\vphi,\vphi^{*}]+\int_{0}^{t}ds\sum_{\alpha}\sigma_{\alpha}(s) n_{\alpha}(s)\right)_{\vphi=\vphi^{\rm mf}}=0 
\end{equation}
satisfying
\begin{equation}
\label{eq:BCs}
\vphi(s=0)=\vphi_{\alpha} {\rm \ \ and \  } \vphi(s=t)=\vphi_{\beta}.
\end{equation}
Finally, the van Vleck-Morette determinant $A(t,[\vec{\sigma}])$ can be shown to be independent of the boundary conditions $\vphi_{\beta},\vphi_{\alpha}$ for quadratic actions, and its precise form can be found in \cite{Schulman81}. 

It is instructive to consider the specific situation for a system described by the Bose-Hubbard Hamiltonian. In this case, besides the interaction term, Eq.~(\ref{eq:Int}), the action functional contains a free term of the form
\begin{equation}
R_{\rm free}[\vphi,\vphi^{*}]=\int_{0}^{t}ds\sum_{j}\Im \left(\phi_{j}(s)\frac{d\phi_{j}^{*}(s)}{ds}\right)+ \int_{0}^{t} ds \sum_{i,j}h_{ij}\phi_{i}^{*}(s)\phi_{j}(s)
\end{equation}
and therefore the mean field solutions that fully determine the reduced propagator for a given auxiliary field $\vec{\sigma}(s)$ are easily obtained from the variation in Eq.~(\ref{eq:EL}) to be
\begin{equation}
\label{eq:ELBH}
i\frac{d}{ds}z_{i}(s)=\sum_{j}h_{ij}z_{j}(s)+\sigma_{i}(s)z_{i}(s),
\end{equation}
namely the mean field equations for a bosonic free field under a time-dependent chemical potential, as expected.

Up to here, the original interacting problem is transformed into a coherent weighted sum of reduced propagators over all possible time-dependent auxiliary fields, each of them representing linear evolution, 
\begin{eqnarray}
\label{eq:Popsig}
K(\vphi_{\beta},\vphi_{\alpha};t)=\int \frac{{\cal D}[\vec{\sigma}]}{Z({\bf v})}&&{\rm e}^{\frac{i}{2}\int_{0}^{t}ds\vec{\sigma}(s){\bf v}^{-1}\vec{\sigma}(s)}A(t,[\vec{\sigma}]) \\ && \times {\rm e}^{i (R_{\rm free}^{\rm mf}(\vphi_{\beta},\vphi_{\alpha},t,[\vec{\sigma}])+\int_{0}^{t}ds\sum_{j}\sigma_{j}(s) n_{j}^{\rm mf}(s,\vphi_{\beta},\vphi_{\alpha},t,[\vec{\sigma}])))}, \nonumber
\end{eqnarray}
which is an exact expression. It is at this stage that the semiclassical approximation, essential for the construction of the control protocol, enters. Evaluating the integral over auxiliary fields in a saddle point approximation, that in turn requires the assumption that the prefactors $A(t,[\sigma])$ are smooth, the full propagator turns out to be dominated by the configurations, $\vec{\sigma}=\vec{\sigma}_{\gamma}$, satisfying the corresponding vanishing of the first variation, namely
\begin{eqnarray}
\frac{\delta}{\delta \vec{\sigma}}\left(\frac{1}{2}\int_{0}^{t}ds\vec{\sigma}(s){\bf v}^{-1}\vec{\sigma}(s) \right.&+&R_{\rm free}^{\rm mf}(\vphi_{\beta},\vphi_{\alpha},t,[\vec{\sigma}]) \\  &+& \left. \left.\int_{0}^{t}ds\sum_{j}\sigma_{j}(s) n_{j}^{\rm mf}(s,\vphi_{\beta},\vphi_{\alpha},t,[\vec{\sigma}])\right)\right|_{\vec{\sigma}_{\gamma}}=0. \nonumber
\end{eqnarray}
Using the implicit dependence of $R_{\rm free}^{\rm mf}$ and $n_{\alpha}^{\rm mf}$ on the auxiliary field as given in Eqs.~(\ref{eq:Rmf}, \ref{eq:nmf}) and the mean-field equations, Eq.~(\ref{eq:EL}), this gives
\begin{equation}
{\bf v}^{-1}\vec{\sigma}_{\gamma}(s)=\vec{n}^{\rm mf}(s,\vphi_{\beta},\vphi_{\alpha},t,[\vec{\sigma}_{\gamma}])
\end{equation}
or
\begin{equation}
\label{eq:mfs}
\sigma_{\gamma,i}(s)=\sum_{j}v_{ij}n_{j}^{\rm mf}(s,\vphi_{\beta},\vphi_{\alpha},t,[\vec{\sigma}_{\gamma}])
\end{equation}
as the set of equations that determines the dominant auxiliary fields $\vec{\sigma}_{\gamma}$. The extreme nonlinearity of this system resulting from the role of $\vec{\sigma}$ as time-dependent contributions to the couplings in the linear equations (\ref{eq:EL}, \ref{eq:ELBH}) makes it natural to expect a countable, but large, number of solutions. 

The implicit problem in Eq.~(\ref{eq:mfs}) can be made more familiar by invoking the dependence of the mean field occupations $n_{\alpha}^{\rm mf}$ on $\vec{\sigma}$ as given precisely by the mean-field equations (\ref{eq:EL}). Substitution of Eq.~(\ref{eq:mfs}) into the equations obtained from Eq.~(\ref{eq:EL}) then results in the explicit solution
\begin{equation}
\label{eq:mfss}
\sigma_{\gamma,i}(s)=\sum_{j}v_{ij}|\phi_{\gamma,j}^{\rm clas}|^{2}
\end{equation}
where $\vphi^{\rm~class}_{\gamma}$ solves the nonlinear equation
\begin{equation}
\label{eq:EL2}
i\frac{d}{ds}\phi_{i}(s)=\sum_{j}h_{ij}\phi_{j}(s)+\sum_{j}v_{ij}n_{i}(s) \phi_{j}(s) 
\end{equation}
for the Bose-Hubbard type of action. Note that the fields $\vec{\sigma}_{\gamma}$ inherit the boundary conditions, Eq.~(\ref{eq:BCs}), finally expressing the equivalence of the semiclassical approximation at the mean field level with a nonlinear classical field equation. This is reassuring as it is expected that the introduction of auxiliary fields at an intermediate step should not change the classical limit of the original interacting problem, precisely given by Eq.~(\ref{eq:EL2}).

The importance of this representation is, however, clear upon return to the semiclassical evaluation of the exact integral in Eq.~(\ref{eq:Popsig}), given now within the saddle point approximation by
\begin{eqnarray}
\label{eq:Popsigsem}
K^{\rm sc}(\vphi_{\beta},\vphi_{\alpha};t)&=&\sum_{\gamma} \frac{{\cal D}_{\gamma}(\vphi_{\beta},\vphi_{\alpha},t)}{Z({\bf v})} \left.{\rm e}^{\frac{i}{2}\int_{0}^{t}ds\vec{\sigma}(s){\bf v}^{-1}\vec{\sigma}(s)}\right|_{\vec{\sigma}=\vec{\sigma}_{\gamma}} \\ &\times& \left.A(t,[\vec{\sigma}]){\rm e}^{i (R_{\rm free}^{\rm mf}(\vphi_{\beta},\vphi_{\alpha},t,[\vec{\sigma}])+\int_{0}^{t}ds\sum_{j}\sigma_{j}(s) n_{j}^{\rm mf}(s,\vphi_{\beta},\vphi_{\alpha},t,[\vec{\sigma}])))}\right|_{\vec{\sigma}=\vec{\sigma}_{\gamma}}, \nonumber
\end{eqnarray}
where the prefactors ${\cal D}_{\gamma}$ result from the usual integration over Gaussian fluctuations around the classical configurations $\vec{\sigma}_{\gamma}$. The second line of Eq.~(\ref{eq:Popsigsem}) contains the reduced propagator for the quadratic action, Eqs.~(\ref{eq:Ksigma}, \ref{eq:Ksex}) in its semiclassical (but exact) form. Thus,
\begin{equation}
\label{eq:fin}
K^{\rm sc}(\vphi_{\beta},\vphi_{\alpha};t)=\sum_{\gamma} \frac{{\cal D}_{\gamma}(\vphi_{\beta},\vphi_{\alpha},t)}{Z({\bf v})} {\rm e}^{\frac{i}{2}\int_{0}^{t}ds\vec{\sigma}_{\gamma}(s){\bf v}^{-1}\vec{\sigma}_{\gamma}(s)} \times K_{\vec{\sigma}_{\gamma}} (\vphi_{\beta},\vphi_{\alpha};t) 
\end{equation}
where $K_{\vec{\sigma}_{\gamma}}$ is the exact coherent state propagator for the $\gamma$-th classical auxiliary field that, given the quadratic nature of the corresponding action, acts invariantly on the initial coherent state, i.e., it transports $\vphi_{\alpha}$ without dispersion along the classical phase space trajectory.  

Equation (\ref{eq:fin}) is the final result of this analysis. It expresses the semiclassical approximation to the coherent state propagator of the interacting theory as a coherent sum over a countable set of contributions. Each term in the sum represents the exact and dispersionless quantum propagation of the initial coherent state under auxiliary fields playing the role of time-dependent chemical potentials that are in turn determined by the solution of the interacting problem. In the language of the main text, the full interacting propagator is given as a sum over all possible classical (mean field) control protocols. The control protocol proposed in the main text amounts then to choosing, based on minimal time $t$ to dynamically connect $\vphi_{\alpha}$ with $\vphi_{\beta}$ or other physical considerations, a single one of the auxiliary fields.

\section{Coherent Time-Evolution in Control Systems}
\label{sec:A2}
A proof is outlined showing that a coherent state $\coh{\Vec{\phi}(t)}$ centered on any mean-field solution $\Vec{\phi}(t)$ of the control Hamiltonian Eq.~\ref{eq:conBH_H} solves the controlled quantum time evolution
\begin{align*}
\coh{\Vec{\phi}(t)}  = \hat{U}_{\rm c}(t) \coh{\Vec{\phi}_{0}}  =  \hat{\mathcal{T}}\exp\Big\{-i\int_{0}^{t} \hat{H}_{\rm c} (s)\d s\Big\} \coh{\Vec{\phi}_{0}},
\end{align*}
where $\Vec{\phi}_{0} = \Vec{\phi}(0)$. More explicitly, the coherent state is shown to be a solution to the corresponding Schrödinger equation, i.e.,
\begin{align*}
i\frac{\d }{\d t} \coh{\Vec{\phi}(t)}  = \hat{H}_{\rm c}(t)  \coh{\Vec{\phi}(t)}\ , 
\end{align*}
which is equivalent to the previous equation. To this end, recall that the classical control Hamiltonian, Eq.~\ref{eq:clBH_Hcont}, can also be written in terms of complex mean-fields
\begin{equation*}
\mathcal{H}_{\text{c}}(\Vec{\phi}, \Vec{\phi}^*, t) = -\sum\limits_{j=1}^L \left\{\phi_{j+1}^* + \phi_{j-1}^* - \mu_j(t)\phi_{j}^*\right\}\phi_{j},
\end{equation*}
that evolve according to the equations motion
\begin{align*}
i\dot{\phi}_j &= \frac{\partial \mathcal{H}_{\text{c}}}{\partial \phi_j^*} = - \left\{\phi_{j+1} + \phi_{j-1} - \mu_j(t)\phi_{j}\right\}\\
i\dot{\phi}_j^* &= -\frac{\partial \mathcal{H}_{\text{c}}}{\partial \phi_j} = \left\{\phi_{j+1}^* + \phi_{j-1}^* - \mu_j(t)\phi_{j}^*\right\}.
\end{align*}

Start by assuming that there is a coherent state solution and look for a contradiction or consistency.  Taking the left hand side time derivative of the coherent state reduces to taking the derivative of the mean-field since the mean particle number conservation along any classical solution implies $\lVert{\vphi}\rVert^{2} = N = \text{const.}$
\begin{align*}
i\frac{\d }{\d t} \coh{\Vec{\phi}(t)}  =& i e^{-\lVert{\vphi}\rVert^{2}/2} e^{\vphi(t) \cdot \vad} \left(\sum\limits_{j=1}^L \dot{\phi}_j(t) \hat{a}_{j}^{\dagger} \right)\ket{0}.
\end{align*}
Next use the classical mean-field equations of motion from above
\begin{align*}
i\frac{\d }{\d t} \coh{\Vec{\phi}(t)} &= -i e^{-\lVert{\vphi}\rVert^{2}/2} e^{\vphi(t) \cdot \vad} \sum_{j=1}^{L}\left(\left\{\phi_{j+1}(t)+\phi_{j-1}(t) - \mu_{j}(t) \phi_{j}(t)\right\}\hat{a}_{j}^{\dagger}\right)\ket{0}\\
&= -\sum_{j=1}^{L}\left(\left\{\phi_{j+1}(t)+\phi_{j-1}(t) - \mu_{j}(t) \phi_{j}(t) \right\}\hat{a}_{j}^{\dagger} \right)\coh{\Vec{\phi}(t)}.
\end{align*}

With respect to the right hand side, use the definition given in Eq.~\ref{eq:coherentStateEigenstate} of a coherent state being an eigenstate of the annihilation operator to obtain 
\begin{align*}
\hat{H}_{\rm c}(t) \coh{\Vec{\phi}(t)}  &= -\sum_{j=1}^{L} \left(\left\{\hat{a}_{j+1}^{\dagger} + \hat{a}_{j-1}^{\dagger} + \mu_j(t)\hat{a}_{j}^{\dagger}\right\} \hat{a}_{j} \right) \coh{\Vec{\phi}(t)}\\
&= -\sum_{j=1}^{L} \left(\hat{a}_{j}^{\dagger}\left\{\phi_{j+1}(t)+\phi_{j-1}(t) + \mu_{j}(t) \phi_j(t)\right\}\right)
\coh{\Vec{\phi}(t)},
\end{align*}
which matches exactly the expression coming from the left hand side. 

\paragraph{Remark 1}
The proof for number projected states is analogous using the property
\begin{equation*}
\hat{a}_{j}\proj{\Vec{\phi}(t)} = \phi_{j}|{\Vec{\phi}(t)}\rangle_{\rm proj}^{N-1}.
\end{equation*}

\paragraph{Remark 2}
The control trajectory used to drive the chemical potentials $\left\{\mu_j(t)\right\}$ is a special case since it is a solution to both the classical equations of motion of the control system, as well as the Bose-Hubbard mean-field limit. With that information, the above proof implies that a coherent state initially centered on the control trajectory arrives at the end of the trajectory with perfect fidelity.

\paragraph{Remark 3}
For any specified control Hamiltonian of the form of Eq.~\eqref{eq:conBH_H}, recalling the equation for calculating the stability matrix, every initial mean-field condition or trajectory leads to the exact same time-dependent stability matrix.  Thus, all trajectories have to rotate around the control trajectory and the stability matrix solution is a time-dependent orthogonal matrix.  This implies that for a coherent state initially centered off the control trajectory by $||\delta\Vec{x}_{\text{init}}||$, this distance stays constant. By using the properties of the stability matrix, and Heller's linearized wave packet dynamics~\cite{Heller75}, it would be possible to provide an analytical formula for the fidelity, for example shown in Fig.~\ref{fig:fidelityNoShifts} based solely on the knowledge of the initial state and the control trajectory.

\clearpage

\bibliographystyle{unsrt}
\bibliography{classicalchaos, extreme, furtherones, general_ref, manybody, molecular, quantumchaos, quantumcontrol, rmtmodify}

\end{document}